\documentclass[manuscript]{aastex}

\shorttitle{The Orientation of Galaxies in Galaxy Clusters }
\shortauthors{God{\l}owski, Piwowarska, Panko,  \& Flin }
 
\begin{document}
 
\title{The Orientation of Galaxies in Galaxy Clusters.}
 
\author{W{\l}odzimierz God{\l}owski}
 
\affil{Uniwersytet Opolski, Institute of Physics, ul.  Oleska  48,
45-052 Opole, Poland}
\email{godlowski@uni.opole.pl}
 
\author{Paulina Piwowarska}
\affil{Uniwersytet Opolski, Institute of Physics, ul.  Oleska  48,
45-052 Opole, Poland}
\email{paoletta@interia.pl}
 
\author{Elena Panko}
\affil{ Odessa National University, Astronomical Department, Park Shevchenko 65014, Odessa, Ukraine}
\email{tajgeta@sp.mk.ua}
 
\author{Piotr Flin}
\affil{Jan Kochanowski University, Institute of Physics, ul. Swietokrzyska 15,
25-406 Kielce, Poland}
\email{sfflin@cyf-kr.edu.pl}
 
\begin{abstract}
We present an analysis of the spatial orientations of  galaxies in the 247
optically selected rich Abell clusters, having in the considered area at
least 100 members. We investigated the relation between angles giving
information about galaxy angular momenta and the number of members in each
structure. The position angles of the galaxy major axes, as well as  two angles
describing the spatial orientation  of galaxy plane were tested for isotropy,
by applying three different statistical tests. It is found that the values of
statistics increase with the amount of galaxies' members, which is equivalent
to the existence of the relation between anisotropy  and number of galaxies in
cluster. The search for connection between the galaxies alignments and
Bautz - Morgan morphological types of examined clusters gave weak dependence.
The statistically marginal relation between velocity dispersion and
cluster richness was observed. In addition, it was found that the velocity
dispersion decreases with  Bautz - Morgan type at almost 3$\sigma$ level.
These results show the dependence of alignments with respect to clusters'
richness, which can be regarded as environmental  effect.
\end{abstract}
\keywords{galaxies: clusters: general}
 
\section{Introduction}
 
One of the most important aim of modern extragalactic astronomy and
cosmology is to solve the problem of  structure formation. There are many
theories used to develop scenarios of structure formations
\citep{Peebles69,Zeldovich70,Sunyaew72,Doroshkevich73,Shandarin74,Dekel85,Wesson82,Silk83,Bower05}.
In the commonly accepted $\Lambda$CDM model, the Universe deems to be spatially
flat, as well as homogeneous and isotropic at appropriate scale. However, the
dimension of this scale is changing with the growth of our knowledge
of the Universe. In this model the structure
were formed from the primordial adiabatic, nearly scale invariant Gaussian
random fluctuations \citep{Silk68,Peebles70,Sunyaew70}. This picture is in agreement
with both the numerous numerical simulations \citep{Springel05,w1,w2} and
the observations. The crucial goal is to determine the
discrimination among different models of galaxy formation. An investigation of
the orientation of galaxies in clusters is regarded as a standard test of
theories of galaxy and large scale structure formation. Thus, theories of the
galaxy formation make predictions regarding to the angular momenta of galaxies
\citep{Peebles69,Doroshkevich73,Shandarin74,Silk83,Catelan96,Li98,Lee00,Lee01,Lee02,Navarro04,Trujillio06}.
This parameter is known for certain for only very few galaxies in structures.
Therefore, instead of the angular momenta, the orientation of galaxies
pertinent to each cluster is investigated. In order to acquire this, either
the distribution of galaxy position angles only or the orientation of galaxy
planes were examined.
 
An interesting problem emerges in the case of dependence on the alignment
to the mass of the
structure. \citet{Godlowski05} suggested that alignment of galaxies in
cluster should increase with the number of objects in particular cluster.
There is no clear empirical evidence that galaxy groups and clusters rotate
(see e.g. \citet{Hwang07}) . Thus, it can be accepted that the
total angular momentum of  galaxy structure is mainly connected  with the
galaxies' spins. Moreover, stronger alignment suggests greater total angular
momentum of galactic groups or clusters.
 
The study of galaxy orientation, which substitutes the
investigation of the galaxy's spin distribution, yields different results.
Nevertheless, it is clear that in isolated  Abell clusters of galaxies only
the dominant  brightest cluster members exhibit the sign of alignment
\citep{Flin91,Trevese92,Kim01,Panko09}.
However, very rich galaxy clusters, such  as A754 \citep{Godlowski98},
A14 \citep{Baier03} or A1656 \citep{Djorgovski83,Wu97,Wu98} have shown
non-random distribution of galaxies.
 
\citet{Godlowski05} suggestion that alignment should increase with richness
of the cluster was already confirmed  by  \citet{Aryal07} (based on the
series of paper \citep{Aryal04,Aryal05a,Aryal06}). They analyzed totally
32 clusters of different richness and BM types, founding that alignment
is changing with the richness as well as BM type of the clusters. However,
both \citet{Godlowski05} and \citet{Aryal07} analysis was qualitative only.
Our aim is to test this hypothesis both qualitatively and quantitatively.
Therefore we decided to examine if alignment of galaxies in clusters depends
on the number of their constitutive members and the BM types using statistical
tests.
 
\citet{Plionis03} suggested that galaxy alignment in cluster  depends on the
velocity dispersion of member galaxies. In order to verify this finding we
also analyzed correlation between alignment and velocity dispersion.
 
The present paper  is organized in standard manner. Section 2 describes our
optical data, section 3 presents the statistical  method used in our
analysis. Section 4 is devoted to our results and their discussion, while
section 5  contains   conclusions  which ends the paper.

\section{Observational data}
 
Our paper is based on the analysis of structures taken from PF catalogue
\citep{Panko06}. The structures were extracted from the Muenster Red Sky
Survey (MRSS hereafter) \citep{MRSS03}. MRSS is optical large scale survey
covering an area of 5000 square degrees in the southern hemisphere. After
scanning  217 ESO plates, it gives the information about 5,5 million galaxies.
PF catalogue, like MRSS, is statistically complete till magnitude value
$m=18^m.3$ and it contains structures having at least ten members in a
magnitude range between $m_3$ and $m_3+3$, where $m_3$ is the  magnitude
of the third brightest galaxy located in the considered structure region.
The catalogue contains hints about each structure. Structures were found
involving the Voronoi tessellation technique, which was in detail described
in our previous paper \citep{Panko09}. The galaxy membership was ascribed
to cluster during construction of the PF Catalogue. We are disposing the
information which galaxy belongs to a given structure. The data for each
galaxy member were taken from the MRSS. These includes: the equatorial
coordinates of galaxies($\alpha$, $\delta$), the  diameters of major and
minor axes of the galaxy image ($a$ and $b$ respectively) and the position
angle of the major axis, $p$. In the present paper we selected rich clusters
having at least 100 members and being identified with one of ACO clusters
\citep{ACO}, which gave us the Bautz-Morgan morphological types (BM types).
There are 239 such objects in the PF catalogue. Moreover, 9 objects  can be
identified with two ACO clusters. We have also taken them into
account, which increased our sample to 248 objects. However, we excluded
from our analysis A3822, which potentially has substructures
\citep{Biviano97,Biviano02}. Therefore, our sample has 247 objects.
 
Data dealing to  velocity dispersion of galaxies were taken from the
literature. We found such clues for 97 clusters
\citep{Alonso99,De Propris02,Fadda96,Mazure96,Muriel02,Struble99}.
 
We had two samples of data. In the first sample all galaxies lying in the
region regarded as cluster were taken into account. In the second sample
only galaxies  brighter than  $m_3+3$ were considered. The second sample
should contain purely cluster members.

\section{The method of investigation}

Historically, two main methods for studying galaxy orientation were
proposed.  In the first one \citep{h4} the analysis of the distribution of
the observed position angle  of  the  galactic image major axis was carried
out. In this approach the face-on  and nearly face-on galaxies were excluded
from the analysis. The second  approach allowed us to use the face-on
galaxies as well. This method, based on the de-projection of the galaxy images,
was originally proposed by \citet{Opik70},  applied by \citet{Jaaniste78} and
significantly modified by \citet{f4,g2,g3}. In this method not only the
distribution of galactic position angles $p$ is being analyzed, but also
another important parameter  - the  galaxy's inclination with respect to the
observer's line of sight $i$ is being considered. Two possible orientations
of the galaxy plane were determined, which gave two possible directions
perpendicular to the galaxy plane. It is expected that one of these normals
corresponds to the direction of galactic rotation axis. Any study of galactic
orientation based on the projection of galaxies on the celestial sphere gives
a four-fold ambiguity in the solution for angular momentum. By the reason of
none information connected with the direction of the galaxy spin our analysis
is reduced to only two solutions.
 
The  inclination  angle  was calculated according to the formula:
$cos^2 i=(q^2 -q^2_0 )/(1-q^2_0)$, where observed axial ratio $q=b/a$
and $q_0$  is  "true" axial ratio. Formula mentioned above is valid for
oblate spheroids \citep{Holmberg46}. Because of the lack of information
about morphological types of galaxies in MRSS catalogue we used standard
value $q_0=0.2$.
 
For each galaxy, two angles are  determined: $\delta_D$ - the
angle between the normal to the galaxy plane and the main plane of the
coordinate system, and $\eta$ - the angle
between the projection of this normal onto the main plane and the direction
towards the zero initial meridian.
Using the equatorial coordinate system, the following relations hold between
angles ($\alpha$, $\delta$, $p$) and ($\delta_D$, $\eta$)
 
\begin{equation}
\sin\delta_D  =  -\cos{i}\sin{\delta} \pm \sin{i}\cos{r}\cos{\delta},
\end{equation}
\begin{equation}
\sin\eta  =  (\cos\delta_D)^{-1}[-\cos{i}\cos{\delta}\sin{\alpha} + \sin{i}
(\mp \cos{r}\sin{\delta}\sin{\alpha} \pm \sin{r}\cos{\alpha})],
\end{equation}
where $r=p-\pi/2$.
 
As a result of the reduction of our analysis into two solutions it is necessary
to consider the sign of the expression:
$S=-\cos{i}\cos{\delta} \mp \sin{i}\cos{r}\sin{\delta}$
and for $S\ge 0$ reverse sign of $\delta_D$, respectively.
\footnote{See \citep{f4} for detailed explanation, however please note that
there is a printed error in formulae for $S$.}  Separately we repeat all
calculations using supergalactic coordinate system \citep{h4}.
 
In order to detect non-random efects in the distribution of the investigated
angles: $\delta_D$, $\eta$ and $p$ we carried out three different statistical
tests.
At first, we checked whether  the distributions of the investigated angles
($\delta_D$, $\eta$ and $p$) in each individual cluster  were isotropic.
In analyzing of the distribution of the two angles $\delta_D$, $\eta$ it is
possible to use galaxies of any orientation - including face-on galaxies.
Therefore for  analysis of the distribution of the angles
$\delta_D$, and $\eta$   all galaxies' members  were considered.
 It is very difficult to determine in a precise manner the position
angles for galaxies seen face-on and nearly face-on.
Moreover, these angles can yield reliable information with respect
to galaxy planes only for galaxies seen edge-on. Thus, in the study
of the position angles distribution, face -on and nearly face-on galaxies
were excluded from the analysis. In this case  galaxy's members with axial
ratio $b/a \le 0.75$  were taken into consideration only.
 
In all applied statistical tests the entire range of the investigated $\theta$
angle  (where for $\theta$ one can put $\delta_D+\pi/2$, $\eta$ or $p$) is
divided into $n$ bins. We use $n=36$ bins of equal width. We repeated  our
analyses using also different values of bin's width; no significant difference
appeared among results.
The $\chi^2$ test yields the critical value of $49.8$ for
$35$ degrees of freedom (at the significance level $\alpha=0.05)$.
 
If deviation from isotropy is a slowly varying function of the $\theta$
angle one can use the Fourier test \citep{h4}:
 
\begin{equation}
N_k = N_{0,k} (1+\Delta_{11} \cos{2 \theta_k} +\Delta_{21} \sin{2\theta_k})
\end{equation}
where  $N_k$ - the number of galaxies within k-th angular bin
and as $N_{0,k}$ - the expected number of galaxies per bin.
 
If theoretical probability function $p_k$ is uniform (i.e $N_{0,k}$ are equal,
as it is in the cases of the $\eta$ and $p$ angles) or symmetric with respect
to the value $\theta=\pi/2$ (i.e. with respect to  value $\delta_D=0$ in the
case of $\delta_D$ angle) we obtain the following expressions for the
$\Delta_{i1}$ coefficients:
 
\begin{equation}
\Delta_{11} = {\sum_{k = 1}^n (N_k -N_{0,k})\cos{2 \theta_k} \over
\sum_{k = 1}^n N_{0,k} \cos^2{2 \theta_k}}, \qquad
\Delta_{21} = { \sum_{k = 1}^n (N_k-N_{0,k})\sin{2 \theta_k} \over
\sum_{k = 1}^n N_{0,k} \sin^2{2 \theta_k}}
\end{equation}
with the standard deviation given by the expressions:
\begin{equation}
\sigma(\Delta_{11}) = \left( {\sum_{k = 1}^n N_{0,k} \cos^2{2 \theta_k}
} \right)^{-1/2} \approx \left( {2 \over n N_0} \right)^{1/2}, \qquad
\sigma(\Delta_{21}) = \left( {\sum_{k = 1}^n N_{0,k} \sin^2{2 \theta_k}
} \right)^{-1/2} \approx \left( {2 \over n N_0} \right)^{1/2}.
\end{equation}
 
The probability that the amplitude
\begin{equation}
\Delta_1 = \left( \Delta_{11}^2 + \Delta_{21}^2 \right)^{1/2}
\end{equation}
is greater than a certain chosen value is given by the formula:
\begin{equation}
P(>\Delta_1 ) =
\exp{\left( -{1 \over 2} \left({\Delta_{11}^2\over
\sigma(\Delta_{11}^2)}+{\Delta_{21}^2 \over \sigma(\Delta_{21}^2)}\right)
\right)}
\approx \exp{\left( -{n \over 4} N_0 \Delta_1^2 \right)}
\end{equation}
with standard deviation of the amplitude:
\begin{equation}
\sigma(\Delta_1) \approx \left( {2 \over n N_0} \right)^{1/2}
\end{equation}
In cases of testing $\eta$ and $p$ angles we can put $=$ instead of
$\approx$.
 
This test was originally introduced by \citet{h4} and
substantially modified by \citet{g2,g3}.
In  the paper \citet{g3} the case
with higher Fourier modes taken into account was discussed:
\begin{equation}
N_k = N_{0,k} (1+\Delta_{11} \cos{2 \theta_k} +\Delta_{21} \sin{2
\theta_k}+\Delta_{12} \cos{4 \theta_k}+\Delta_{22} \sin{4\theta_k}+.....).
\end{equation}
Amplitude $\Delta$ it is now the function for all four $\Delta_{ij}$ coefficients.
Generally, when we take into consideration both Fourier modes $2\theta$
and $4\theta$ , the formulas are complicated and this was discussed in details
in  \citet{g3}
\footnote{However, please note that there is a printed error in \citet{g3}.
eq. 18 should have form: $P(\Delta)=(1+J/2)\exp{(-J/2)}$)}.

From the sign  of $\Delta_{11}$ coefficient one can deduce the direction of
departure from isotropy. If $\Delta_{11}<0$, then the excess of the galaxies
with $\theta$ angle near $90^o$ is observed. It indicates for example that in
the case of the position angles ($\theta \equiv p$)
$\Delta_{11}<0$ means that the excess of galaxies with position
angles near $90^o$ (parallel to main plane of the coordinate system)
is observed. If $\Delta_{11}>0$ then the  excess of objects with position
angles perpendicular to the main plane of the coordinate system is observed.
Therefore, for $\Delta_{11}>0$ the rotation axis projections tends to be
parallel to the main plane.
 
In the case of $\delta_D$ angle the situation is more complicated.
However, if we restrict our analysis only to the case of the absolute value of
$\delta_D$ angle, then we could neglect  $\Delta_{21}$ and $\Delta_{22}$
coefficients, because they are equal to zero (see also
\citep{f4,Aryal04,Aryal05a,Aryal06,Aryal07}).
In that case $\Delta_1$ is reduced to $|\Delta_{11}|$, while $\Delta$, now
denoted as $\Delta_c$, is the function of coefficients $\Delta_{11}$ and
$\Delta_{12}$ only. However, please note that $\Delta_{11}$ and $\Delta_{12}$
are not independent of each other \citep{g3}.
 
We also performed the investigation of the linear regression given by  $y=aN+b$
counted for various parameters. These were carried out for each  investigated
angle separately. We studied  the linear regression  between: the values of
different statistics
$\chi^2$, $\Delta_1/\sigma(\Delta_1)$, $\Delta/\sigma(\Delta)$ and the number
of analyzed galaxies in  each particular cluster. In the case of $\delta_D$,
 the values of statistics :
$\chi^2$, $|\Delta_{11}/\sigma(\Delta_{11})|$, $\Delta_c/\sigma(\Delta_c)$
and the number of analyzed galaxies in  each particular cluster were
considered.
 
We assumed that the theoretical, uniform, random distribution contains the
same number of objects as the  observed one. Our null hypothesis $H_0$  is
that the distribution is a random one.  In such a case the  statistics
$t=a/\sigma(a)$ has Student's  distribution with $u-2$ degrees of freedom,
where $u$ is the number of analyzed clusters. It means that we tested $H_0$
hypothesis that $t=0$ against $H_1$ hypothesis that $t>0$. In order to reject
the $H_0$ hypothesis, the value of the observed statistics $t$ should be
greater than $t_{cr}$. Our sample has 247 clusters. For this sample, at the
significance level $\alpha=0.05$,  the value $t_{cr} = 1.651$, while for the
sample of 97 clusters with known values of velocity dispersion, the value
$t_{cr} = 1.660$.
 
Similarly, using linear regression we looked for possible relations between the
values of applied statistics:
$\chi^2$, $\Delta_1/\sigma(\Delta_1)$ and $\Delta/\sigma(\Delta)$ (or
$\chi^2$, $|\Delta_{11}/\sigma(\Delta_{11})|$ and $\Delta_c/\sigma(\Delta_c)$
in the case of $\delta_D$ angle) and the BM type of each cluster. The linear
regression between values of above mentioned statistics and  velocity
dispersion of galaxies inside cluster was also examined.
 
The linear regression analyses were performed independently for the sample
containing all galaxies in the cluster area (sample $A$), as well as for the
sample restricted to galaxies brighter than $m_3+3$ (sample $B$).

\section{Results and discussion}
 
We analyzed the distribution of three angles connected with the orientations of
galaxies in the sample of 247 rich Abell clusters.  The the distribution
of the position angels, $\delta_D$ angles and $\eta$  angles for two
previously investigated clusters  A2721 and A2554 \citep{Aryal07}  were
presented in the Fig.1 Our results are very similar to that obtained
by \citet{Aryal07}  i.e. isotropic distribution for position angles and
spin vectors of galaxies tend to lie in the Local Supercluster plane.
Also our result for $\eta$ angles are similar to \citep{Aryal07}.  Hovewer
our interpretation seems to be a litle bit different - projection of the spin
vectors of galaxies tends to be oriented in direction $L=-45^o$,  not
perpendicular with respect to the Virgo Cluster centre.
 
The dependencies on the
values of the applied statistics used in our study and the cluster richness
for three investigated angles: $p$, $\delta_D$, and $\eta$ were shown on
Fig.2-4 respectively. We studied the linear regression $y=aN+b$ between the
values of each statistics and three parameters, namely: the number of galaxies
in cluster $N$, cluster morphological type $BM$ and galaxy velocity
dispersion $\sigma(V)$. The above calculations were performed separately for
samples $A$ and $B$, as well as for each tested for isotropy angle $p$,
$\delta_D$ and $\eta$. The values of the regression coefficients $a$ and $b$,
with its errors for each investigated case were collected in Table 1
(equatorial coordinate sysem) and Table 2 (supergalactic coordinate sysem).
 
Fig.2  present  the dependence on the applied three  statistics
$\chi^2$, $\Delta_1/\sigma(\Delta_1)$ and $\Delta/\sigma(\Delta)$ for
position angle and cluster richness $N$. Fig.3 and Fig.4 present the same
relations for $\delta_D$ and $\eta$ angles respectively.
For the position angle $p$ only the $\chi^2$ test  and only when we use the
equatorial coordinates as a reference system statistics $t=a/\sigma(a)=1.67$
is greater than $t_{cr} =1.65$ at the significance level $\alpha=0.05$.
In equatorial cordinates, in the case of the $\delta_D$ angle all applied
statistics showed $t>t_{cr}$.
We obtained $t=3.11$ in the case of $y=\chi^2$ test, $t=2.35$ for
$y=|\Delta_{11}/\sigma(\Delta_{11})|$ and $t=3.26$ for
$y=\Delta_c/\sigma(\Delta_c)$. The distribution of the $\eta$ angle is even
more anisotropic, because we obtained  $t=6.0$ in the case of $y=\chi^2$ test
and  $t=3.83$ and $t=5.50$  in two remaining tests respectively.
In supergalactic coordinates the picture is similar but the efect is much
stronger than in the case of the equatorial coordinates.
For restricted sample to galaxies brighter than $m_3+3$,  anisotropy
is even stronger when for the sample including all galaxies in the cluster area,
as well as for the position angles even in the case of the Fourier test statistics
$t=a/\sigma(a)$ reached critical value $t_{cr} =1.65$. This results clearly
confirm \citet{Godlowski05} suggestion that the alignment of galaxies in
clusters increased with the number of members in  clusters.
 
 We have found only weak correlation and only in the case of using
supergalactic coordinates,  between values of the analyzed statistics and
$BM$ type of the clusters. We found a weak correlation between the values of
the analyzed statistics and velocity dispersion  $\sigma(V)$ for $p$ angle and
$\delta_D$ angle. However, only for position angle and only in the case
of Fourier test the value of the statistics $t=a/\sigma(a)$ reached the
critical value $t_{cr} =1.66$.  Moreover, for supergalactic coordinates, we
obtain negative dependence for angle $\delta_D$  in the case of the Fourier test.
 
We also investigated the relation between the number of galaxies belonging
to the cluster and $BM$ type, the velocity dispersion of galaxies belonging
to the cluster and cluster $BM$ type, as well as between cluster richness
and velocity dispersion (Fig5, Tab.3). In Fig.5a we present the
relation between the  cluster richness and $BM$ type. The values of
statistics are smaller than $t_{cr}$. It means, that we do not observe
the relation between richness and $BM$ type. Fig.5b shows the relation
between velocity dispersion $\sigma(V)$ and $BM$ type. In this case  the
value of statistics $t=2.99$, which is greater than $t_{cr}$, therefore we
conclude that  velocity dispersion decreases with $BM$ type. This effect is
almost at  3$\sigma$ level. In Fig. 5c the dependence on the cluster richness
and the velocity dispersion is presented. The value $t=a/\sigma(a)= 1.31$ in
the case of $p$ angle and $t=1.18$, when considering the orientation of galaxy
planes.
 
\section{Conclusions}
 
We analyzed the alignment of galaxies belonging to  247 Abell clusters
containing at least 100 members. Using statistical tests we confirmed
suggestion of \citet{Godlowski05} that non randomness of galaxy orientation
in clusters increases significantly with the cluster  richness. Such
confirmation follows from the analyses of all three investigated  angles
$\delta_D$, $\eta$ and $p$. These angles are connected with the orientation
of galaxies and therefore with the distribution of galaxy angular momenta.
The effect increases if we restricted the cluster membership to galaxies
brighter than $m_3+3$, which suggest that this effect is really connected
with clusters.
 
The observed dependency on the alignment of galaxies in clusters and
richness of the cluster leads to the conclusion that angular momentum of the
cluster increases with the mass of the structure. Usually this dependence is
presented as empirical relation $J\sim M^{5/3}$
\citep{Wesson79,Wesson83,Carrasco82,Brosche86}.
The aim of this relation has been discussed for a long time. One of the first
explanation was proposed by \citet{Muradyan75} in terms of the Ambarzumian's
superdense cosmogony, while  \citet{Mackrossan87} involved termodynamical
consideration  for its explanation. \citet{Wesson83} argued that this is a
consequence of self similarity of Newtonian problem applied to rotating
gravitationally bound systems. The relation was used for pointing out
its possible role in the unification of the gravitation and particle physics
\citep{Wesson81}. \citet{Catelan96} joined the relation $J\sim M^{5/3}$ with
the model of galaxy formation. They explained it as a consequence of the
tidial torque model. \citet{Sistero83} incorporated the rotational velocity
of the Universe. A similar approach was presented by \citet{Carrasco82} who
explained this relation as a consequence of mechanical equilibrium
between the gravitational and rotational energy, while \citet{Li98} proposed
more general relation for  $J(M)$ and explained it as a result of the influence
of the global rotation of the Universe on the structure formation. Our finding
is in agreement with prediction of the Li model \citep{Li98,Godlowski05}.
 
In our opinion the observed relation between the richness of galaxy cluster and
 the alignment is due to tidal torque, as suggested by \citet{Catelan96}.
Moreover, the analysis of the linear tidal torque theory
is pointing in the same direction \citep{Noh06a,Noh06b}. They noticed the
connection of the alignment with the considered scale of structure.
 
We found a strong correlation between $BM$ type and the velocity dispersion.
The velocity dispersion decreases with $BM$ type.  We found only
weak correlation between the alignment and $BM$ type, claimed by
Aryal and Saurer \citep{Aryal04,Aryal05a,Aryal06,Aryal07}.  Our sample of
clusters is  an order of magnitude greater than that one  analyzed by them.
Moreover this weak correlation is found only in the case of using
supergalactic coordinate system as the reference system.
The correlation between the alignment and  velocity dispersion of galaxies
belonging to clusters was found by \citet{Plionis03}. In our data this
effect is statistically  insignificant (it is at $1,5\sigma$ level).
Moreover, it is noted only in the case of $A$ sample, not in $B$ sample
restricted to galaxies brighter than  $m_3+3$. In  PF Catalogue
the position angles of brighter galaxies, which is the brightest member,
the second, third and tenth brightest galaxy are distributed  randomly
\citep{Panko09}, while the present analysis of all galaxies in cluster shows
an anisotropic distribution.  If brighter galaxies are located more centrally,
and dimmer ones are located outside rich clusters
somewhere in the sheets and filaments in which clusters are embedded, then
the discussed possible effects are  environmental ones. When the
distribution of dimmer galaxies in cluster follows that one for brighter
galaxies, the appearance of the alignment in richer region can also be regarded
as the environmental effect. Therefore, we concluded that the richness -
alignment relation is of environmental origin. The performed analyses have shown
that there are several clear as well as somewhat obscure relations between
such parameters as: the alignment of galaxies in clusters, the velocity
dispersion, cluster $BM$ type and density of galaxies in clusters. Further
investigation in this direction should reveals the connection among these
parameters, which should sheds light to formation and evolution of clusters.

\section*{Acknowledgments}
 
This research has made use of the NASA/IPAC Extragalactic Database (NED)
which is operated by the Jet Propulsion Laboratory, California Institute
of Technology, under contract with the National Aeronautics and Space
Administration.  EP thanks the Jan Kochanowski University for hospitality
and financial support during her stays in Kielce. This work was partially
supported by the  Jan Kochanowski University grant BS 052.
We thanks the anonymous referee for suggestions and  comments which help
improving the paper.

\appendix
 
\section{The list of investivated clusters}
 
A list of the investigated clusters is given in Table 4.
The particular columns give the clues about: $PF Number$- the name of the each
structure in Panko-Flin Catalogue, $\alpha$ and $\delta$ of the cluster center,
$N_0$- the number of galaxies in the field of structure, $N_1$- the number of
galaxies with $b/a \le 0.75$, $Name$- structure identification with \citep{ACO}.
The last three columns gives the galaxies alignments (with respect to supergalactic
coordinate system). We presented the results for supergalactic position angle $P$,
polar angle $\delta_D$ and azimuthal angle $\eta$. Bottom indexes $0$ and $?$
denote isotropic distribution and situation that we can not decided if it's
anisotropic or isotropic distribution, respectively.
For $P$ and $\delta_D$ angles $\parallel$ index denotes that spin vector of
galaxies tends to be parralel to supergalactic plane, while $\perp$ index denotes
that spin vector of galaxies is perpendicular to the supergalactic plane.
For the $\eta$ angle $\parallel$ index denotes that projection of the spin vector
to the supergalactic plane tends to be parralel to the direction connecting
Galaxies with the Virgo Cluster (Supercluster center).  The $\searrow$ index
denotes that in all cases we observed anisotropic distribution but no special
direction of deviation from isotropy (especialy parallel or perpendicular)
was found.

\clearpage
 
\begin{figure}
\epsscale{.30}
\plotone{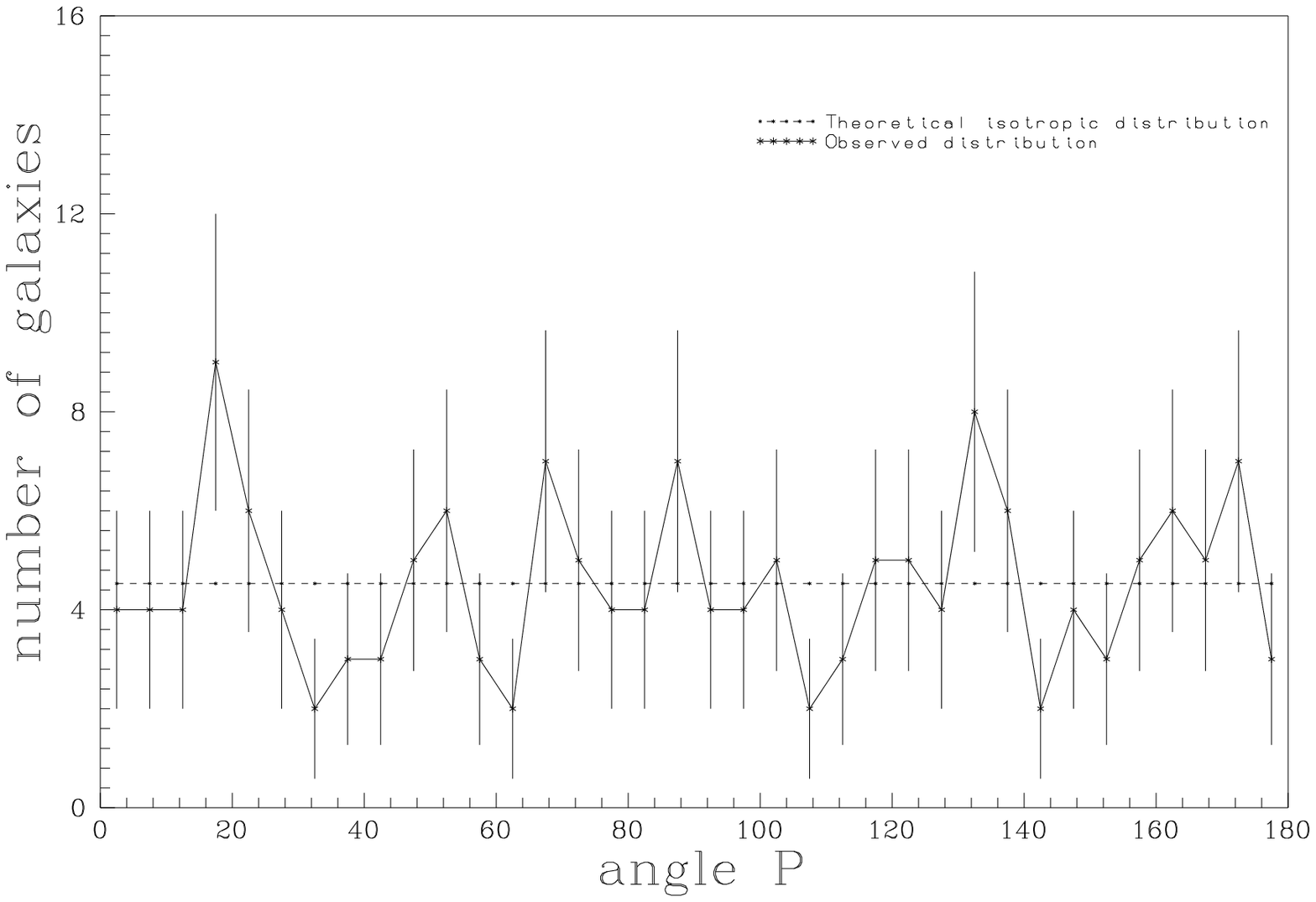}
\plotone{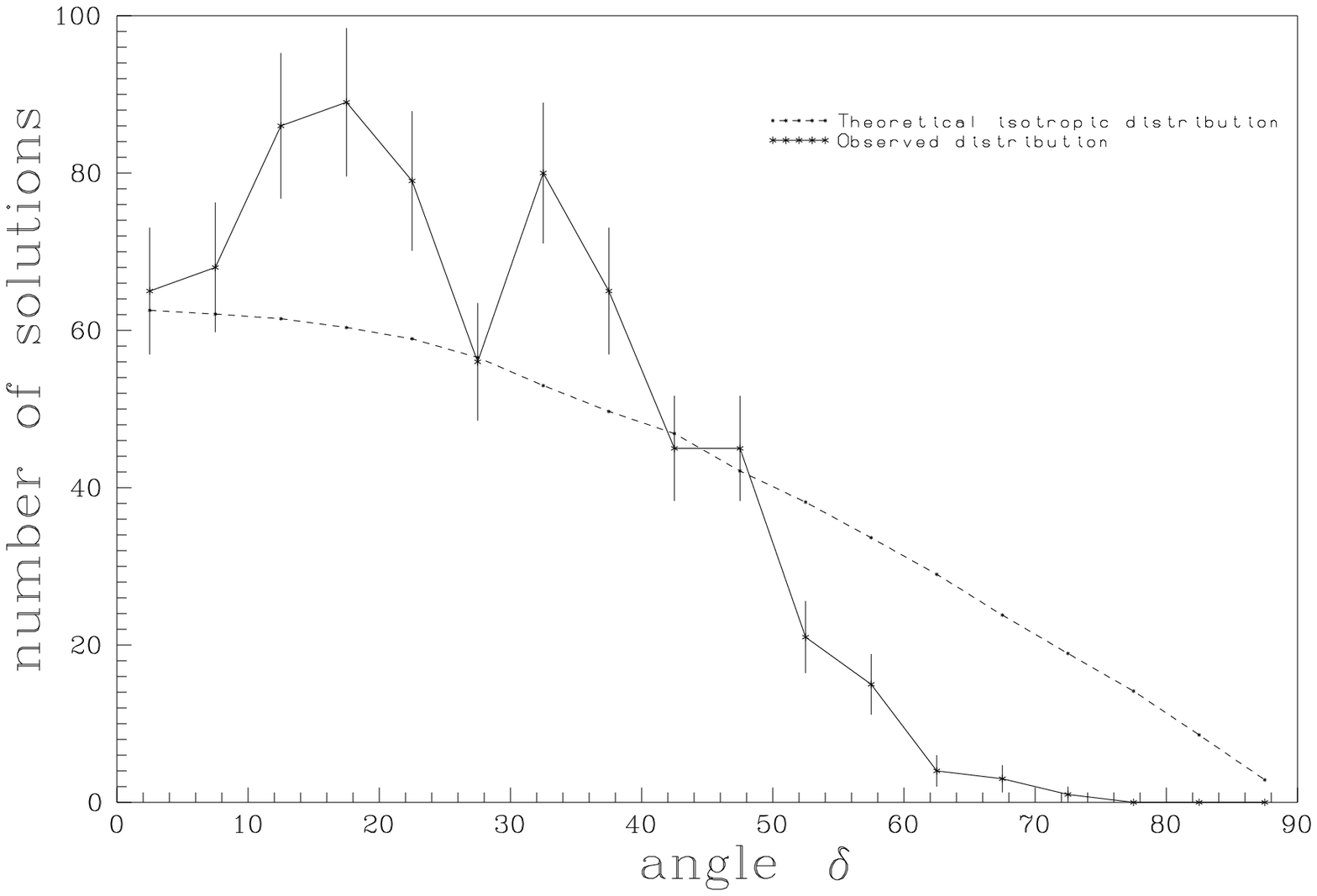}
\plotone{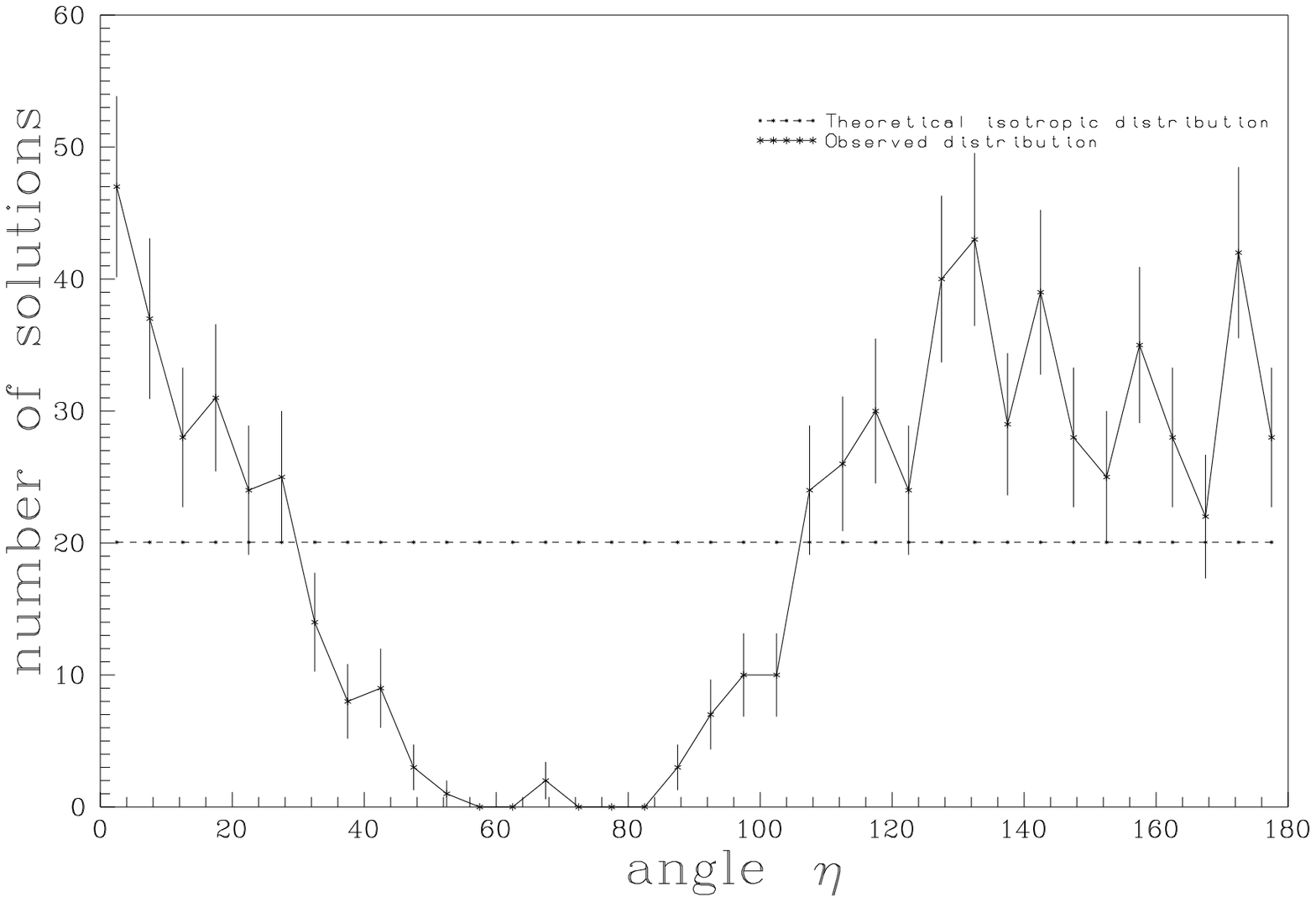}\\
\plotone{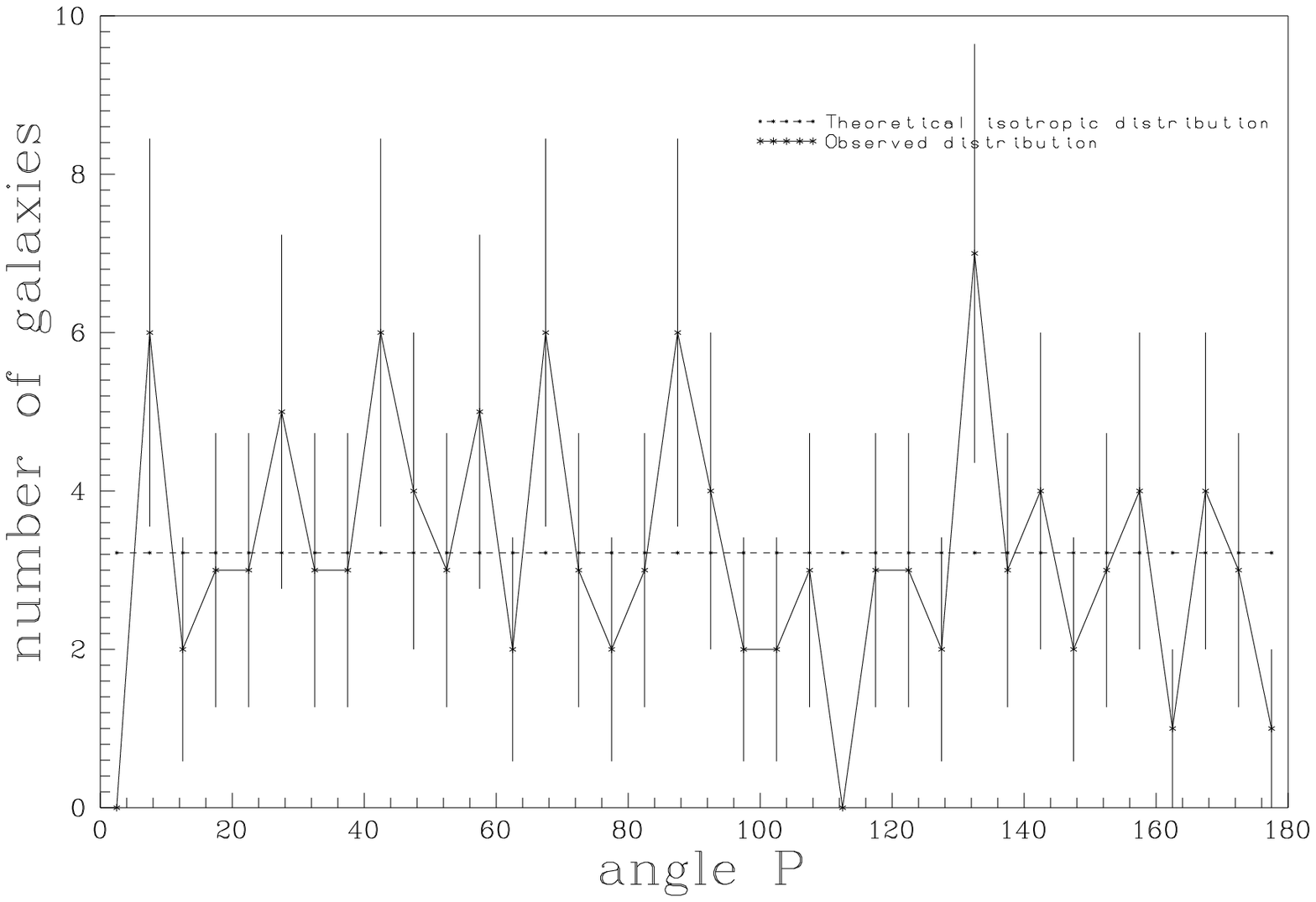}
\plotone{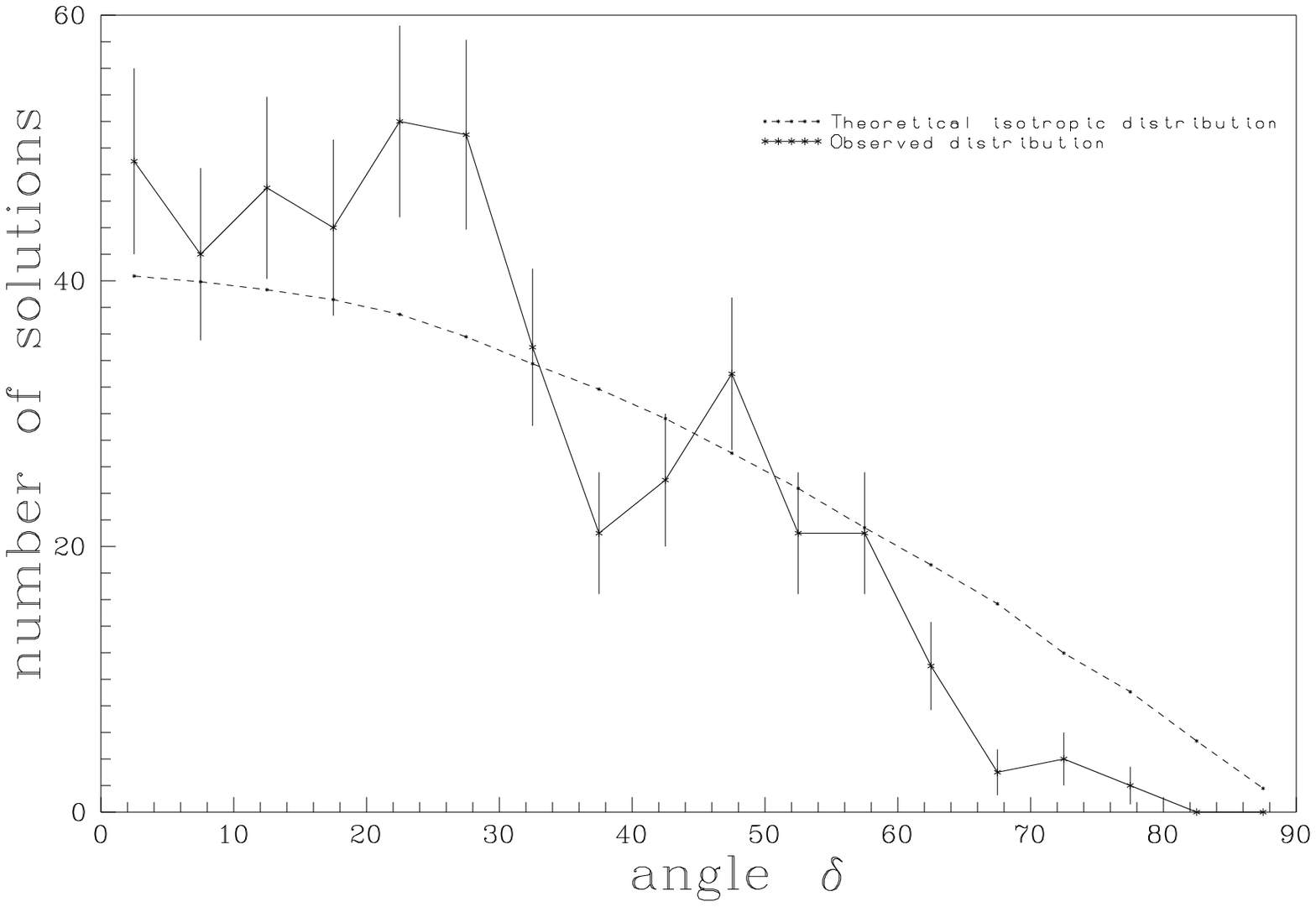}
\plotone{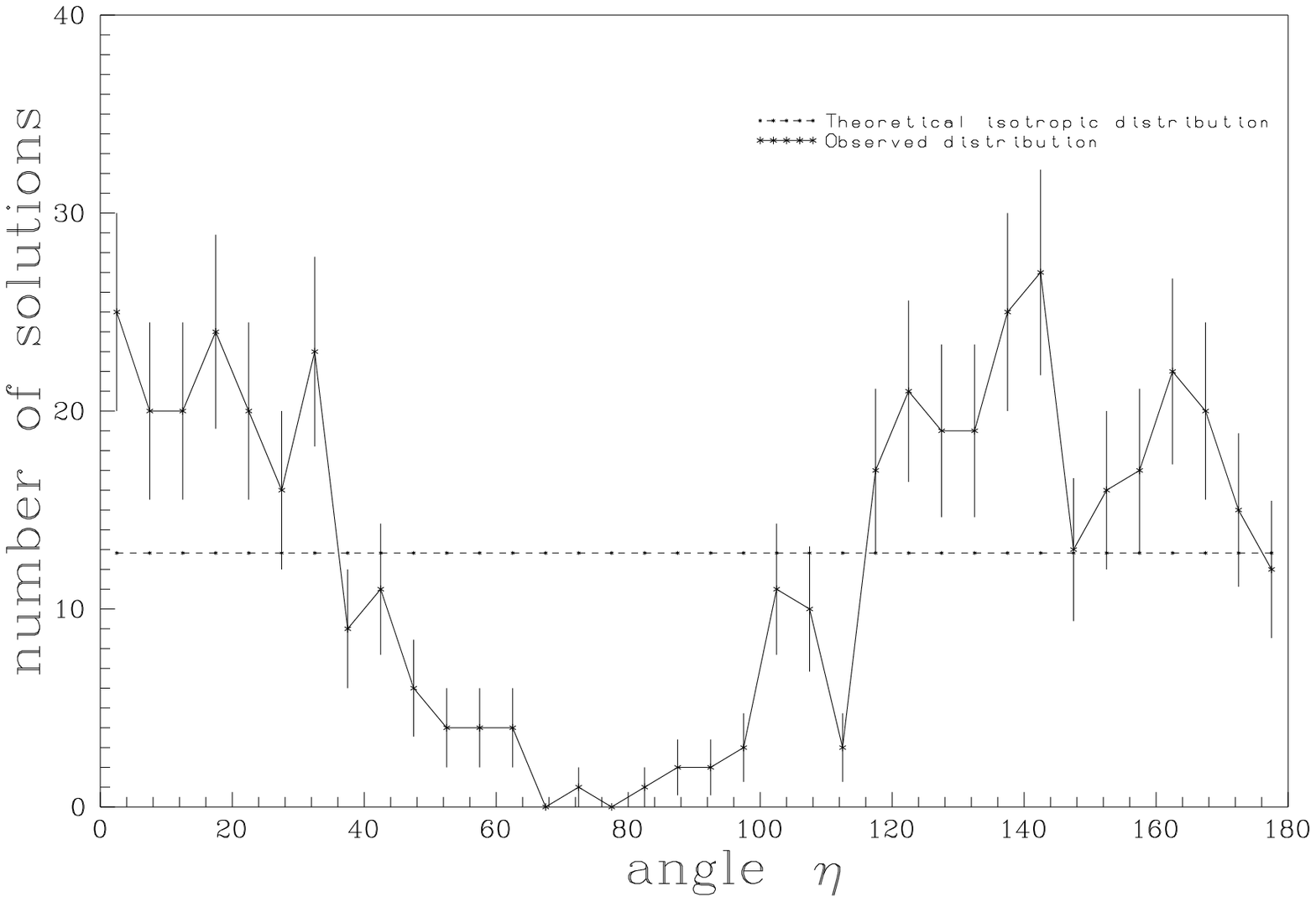}
\caption{The  distribution of the position angels - left panel,
$\delta_D$ angles - middle panel and $\eta$ angles - right panel,
for clusters: A2721 - upper panel and - A2554 bottom panel
(supergalactic coordinate system). We presented theoretical isotropic 
distributions (dashed lines)  and observed distributions. The error 
bar were presented as well.
\label{fig1}}
\end{figure}

\clearpage
 
\begin{figure}
\epsscale{.30}
\plotone{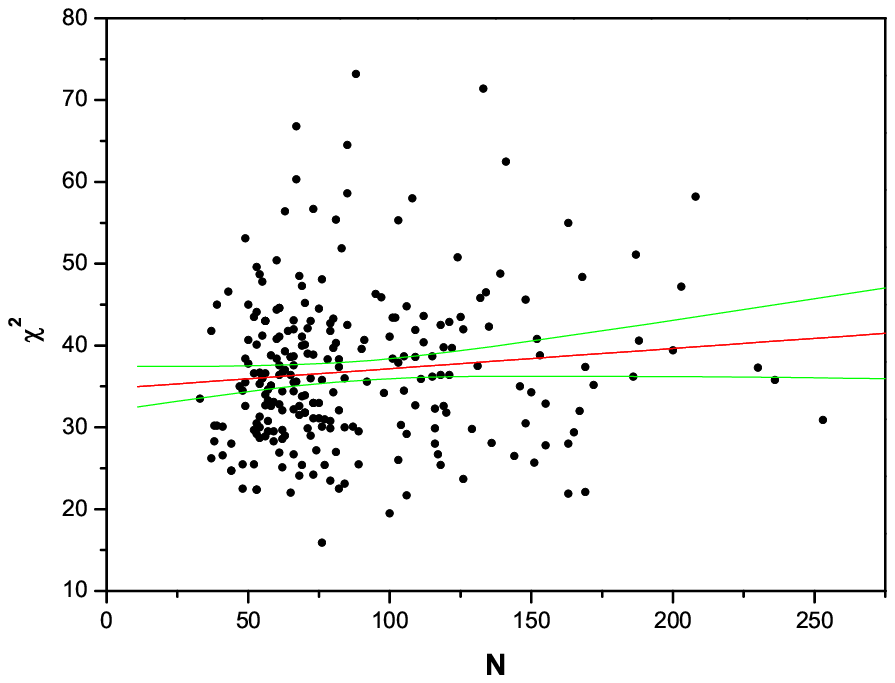}
\plotone{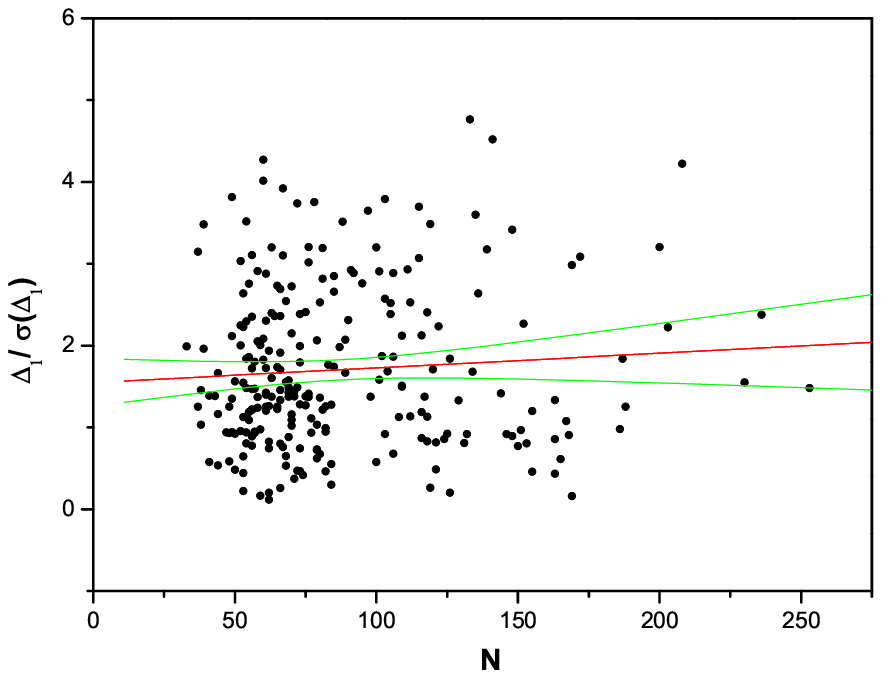}
\plotone{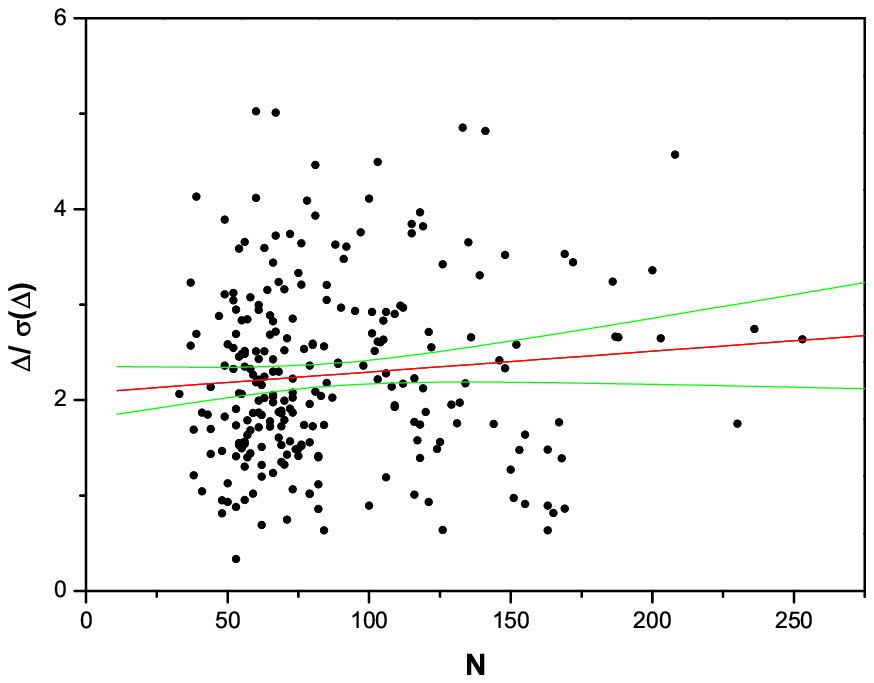}\\
\plotone{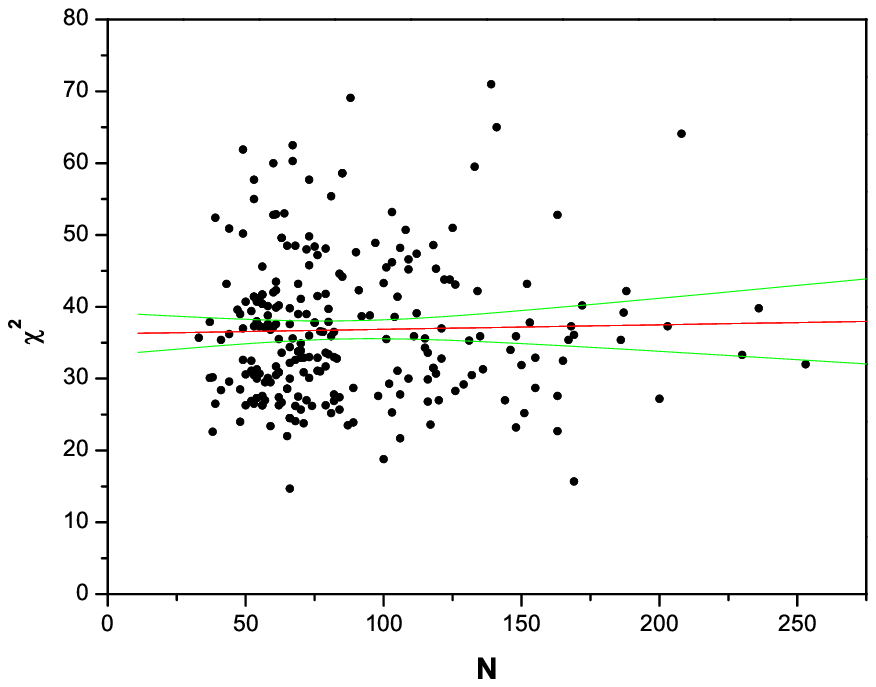}
\plotone{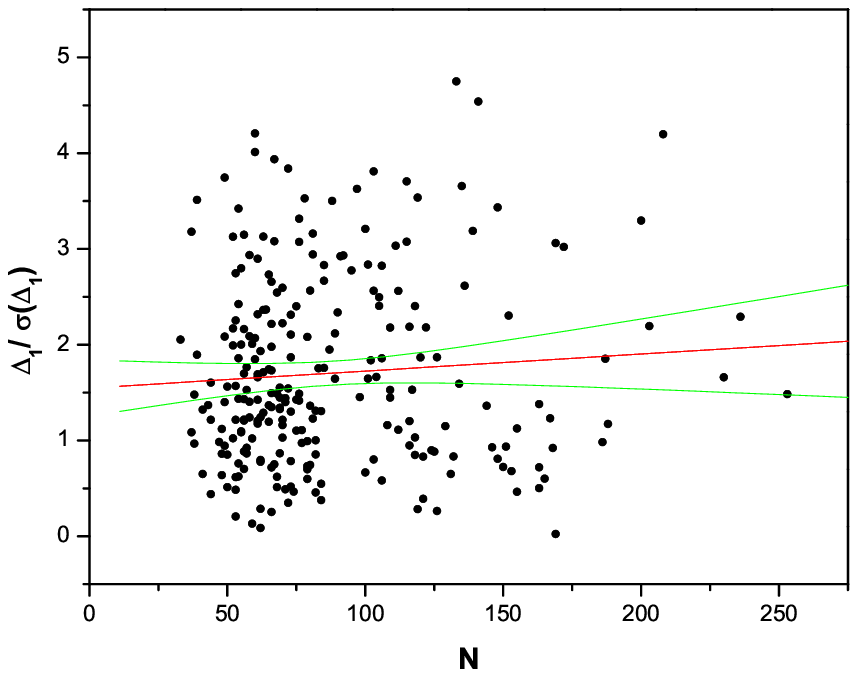}
\plotone{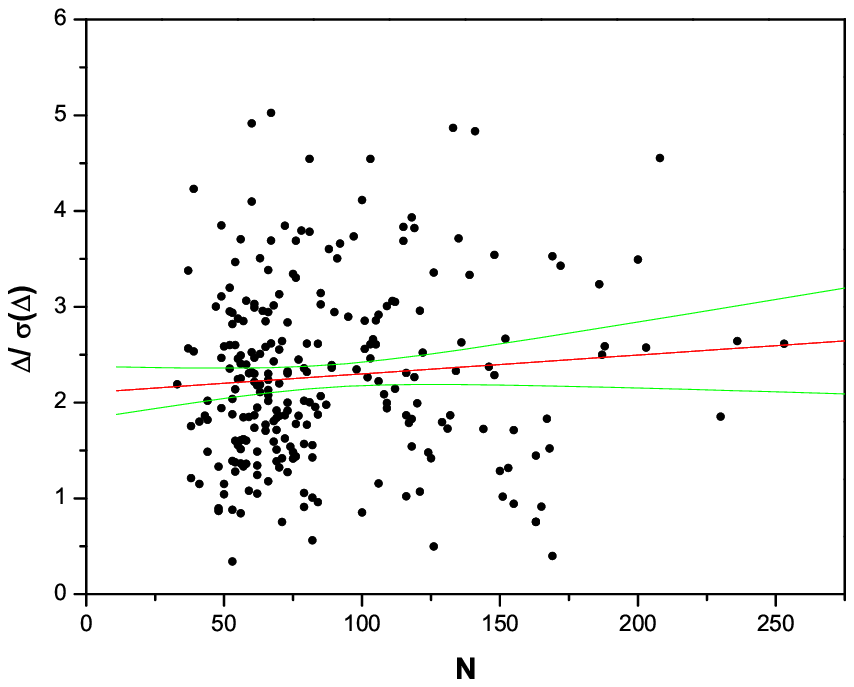}
\caption{The relation  between the number of galaxies in the cluster $N$
and the value of analyzed statistics
($\chi^2$ - left panel, $\Delta_1/\sigma(\Delta_1)$ - middle panel,
$\Delta/\sigma(\Delta)$ - right panel)  for the position angles $p$.
Upper panel - equatorial coordinates, bottom panel - supergalactic coordinates.
The bounds error, at confidence level $95\%$, were presented as well.
\label{fig2}}
\end{figure}
 
\clearpage
 
\begin{figure}
\epsscale{.30}
\plotone{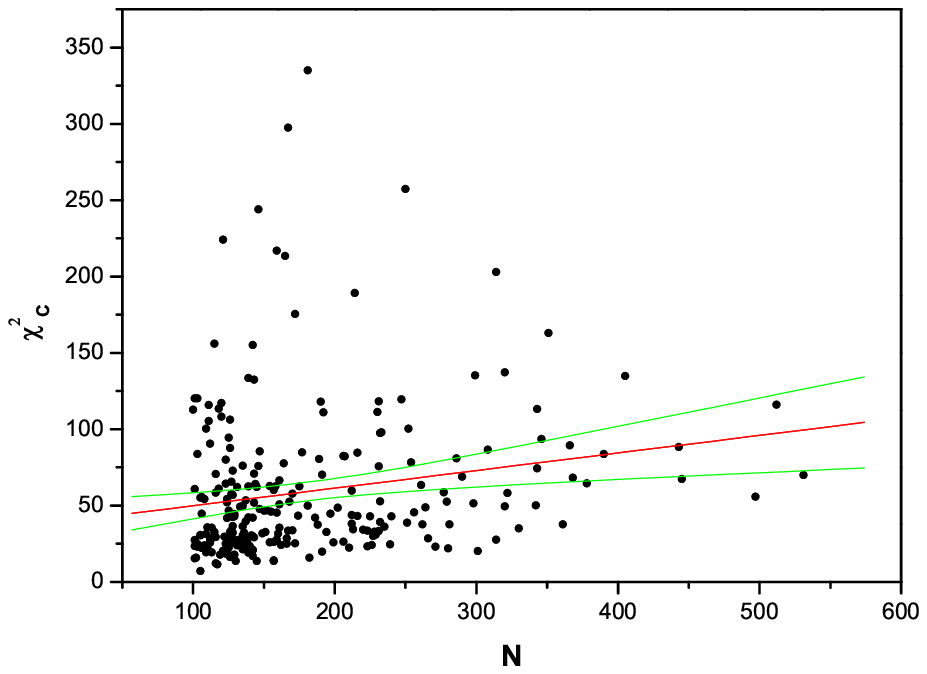}
\plotone{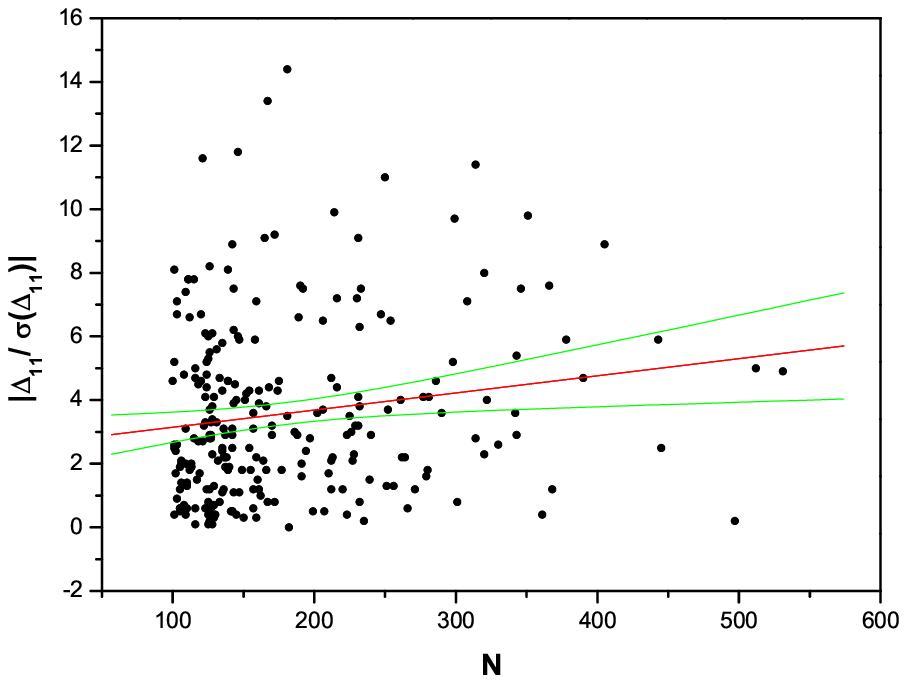}
\plotone{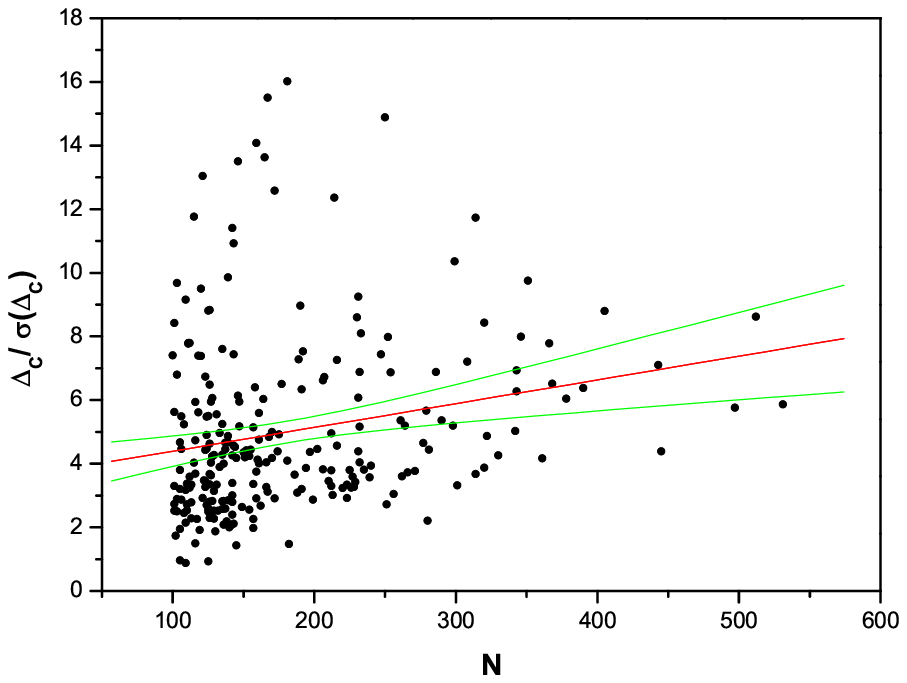}\\
\plotone{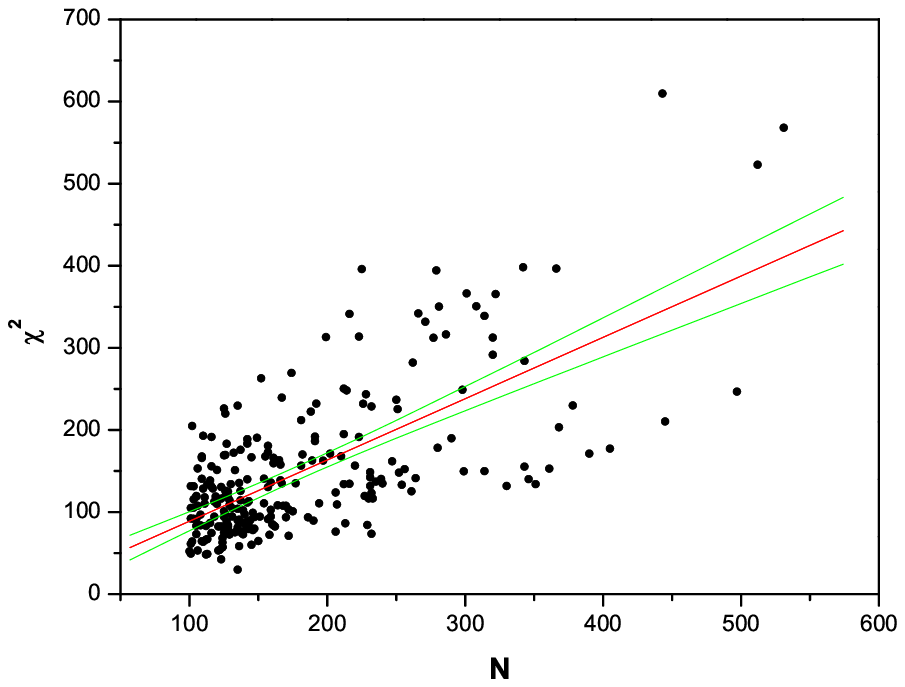}
\plotone{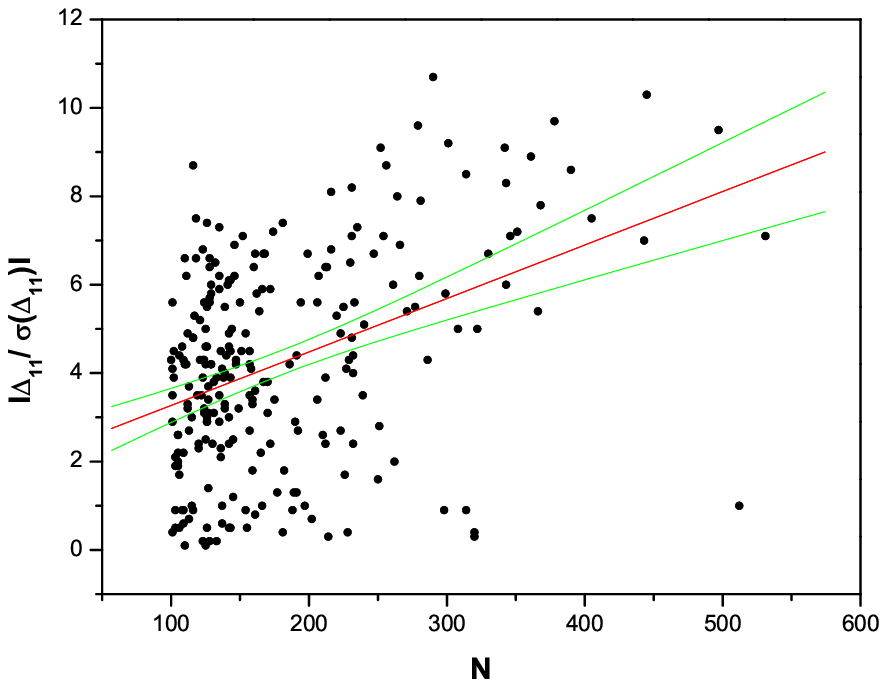}
\plotone{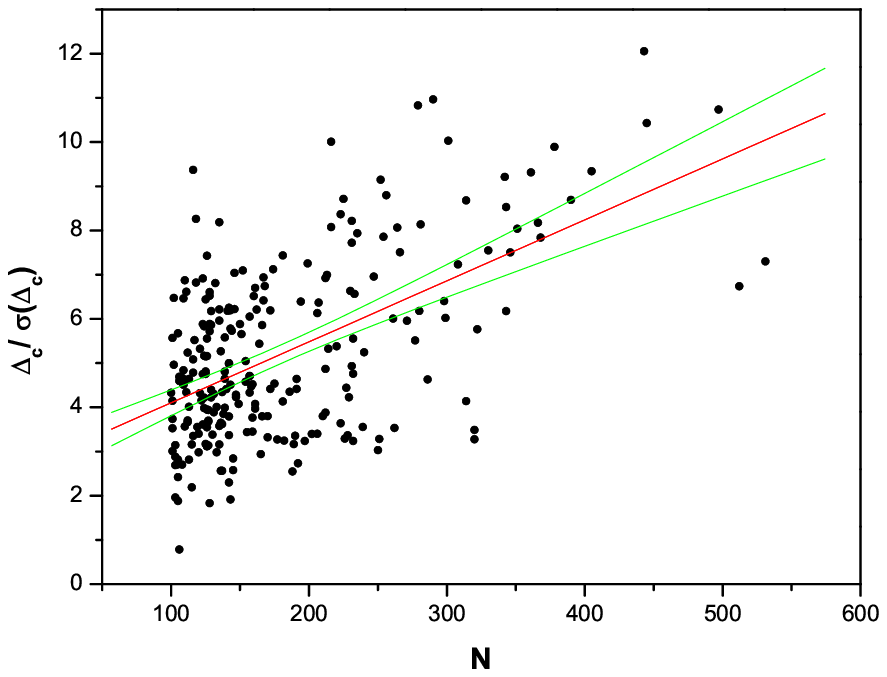}
\caption{The same as in fig.2 but for $\delta_D$ angles.
Upper panel - equatorial coordinates, bottom panel - supergalactic coordinates.
\label{fig3}}
\end{figure}
 
\clearpage
 
\begin{figure}
\epsscale{.30}
\plotone{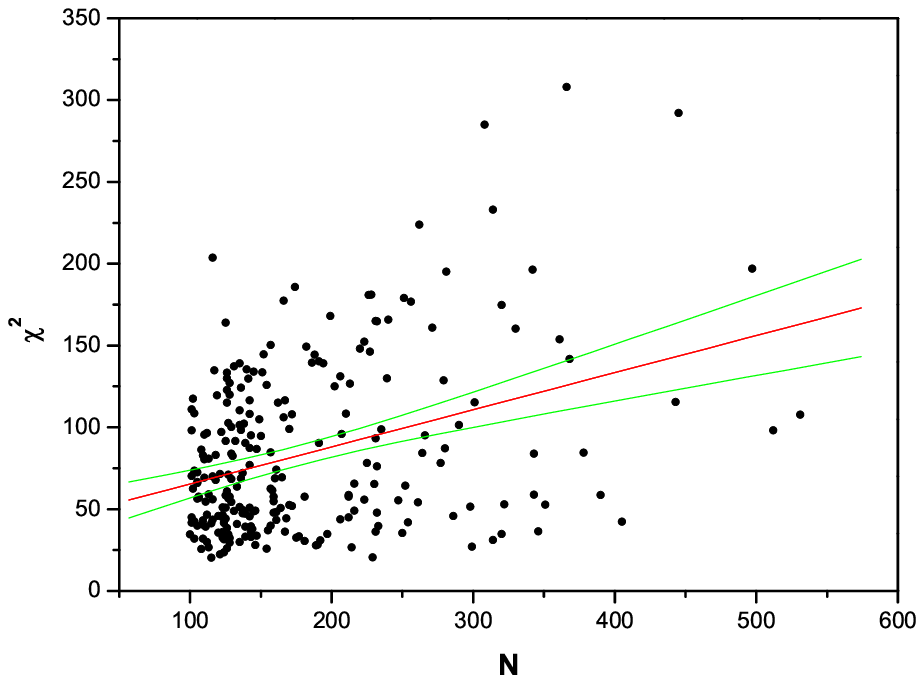}
\plotone{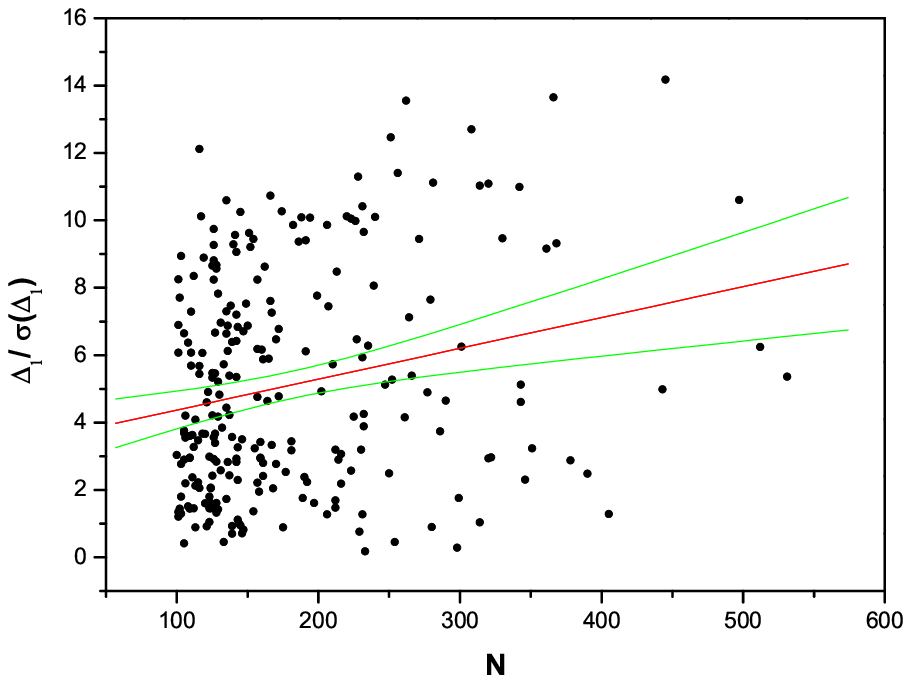}
\plotone{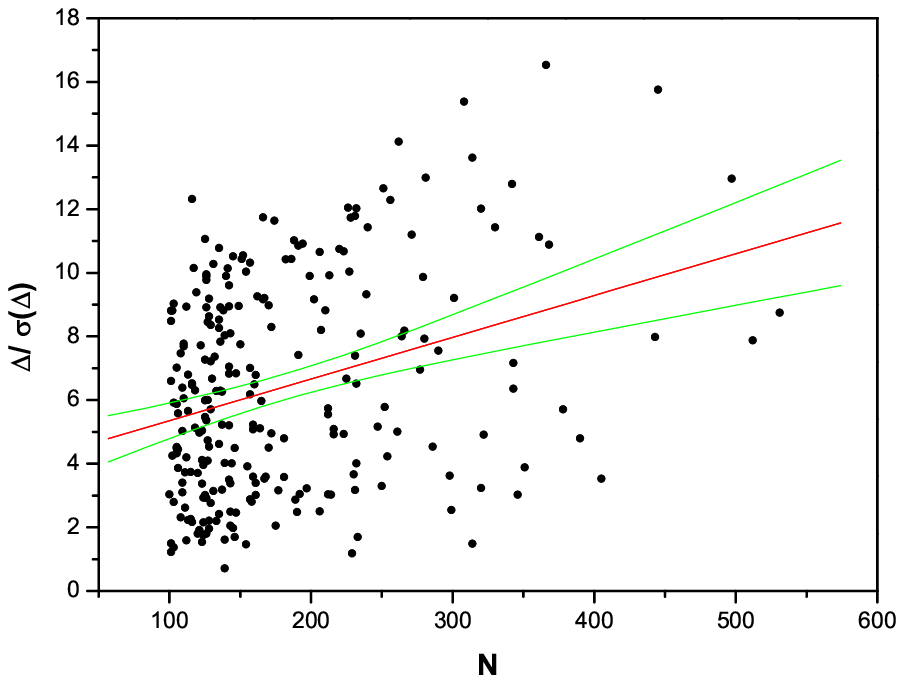}\\
\plotone{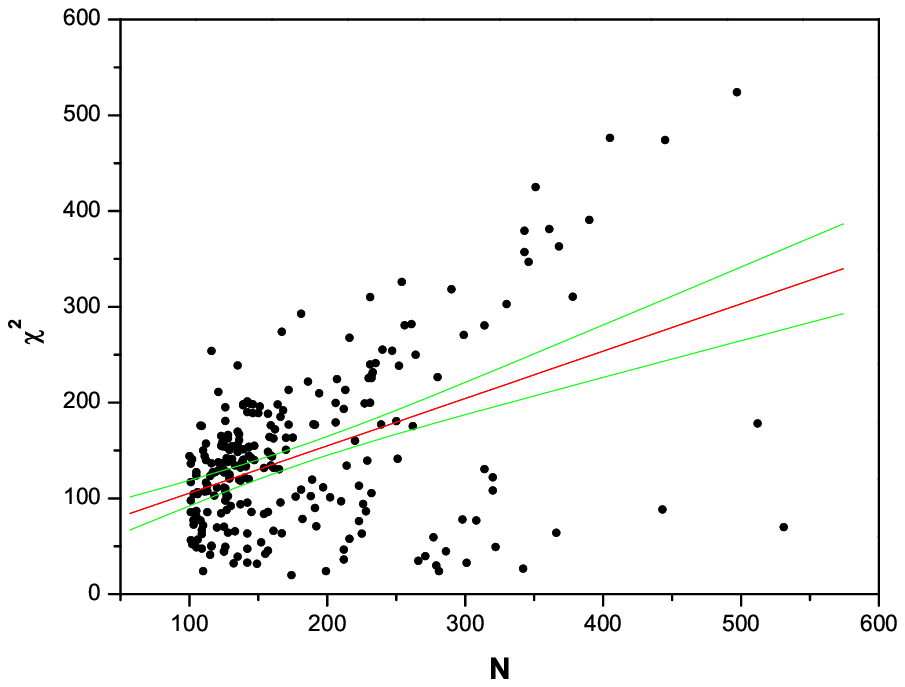}
\plotone{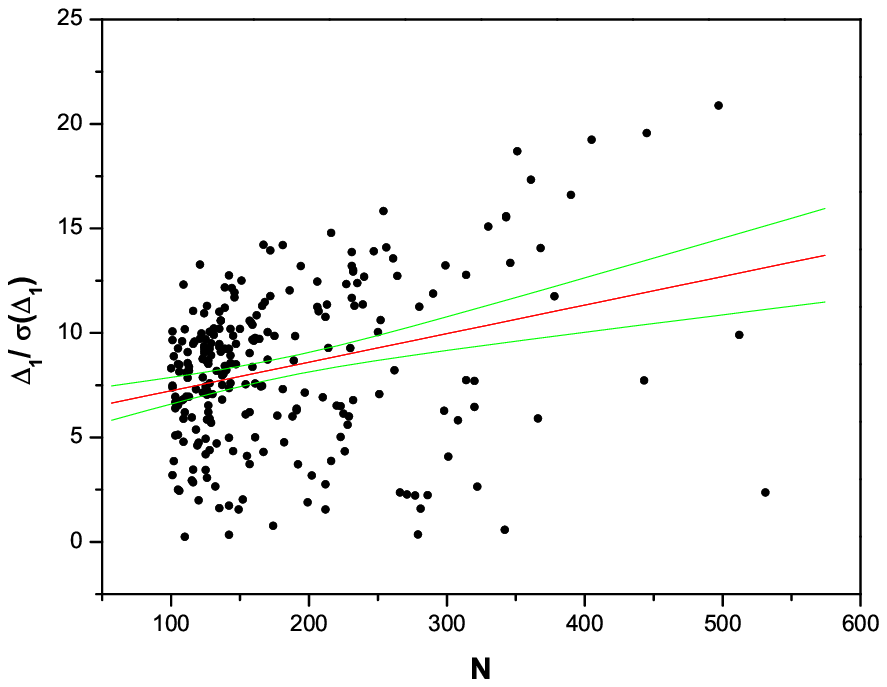}
\plotone{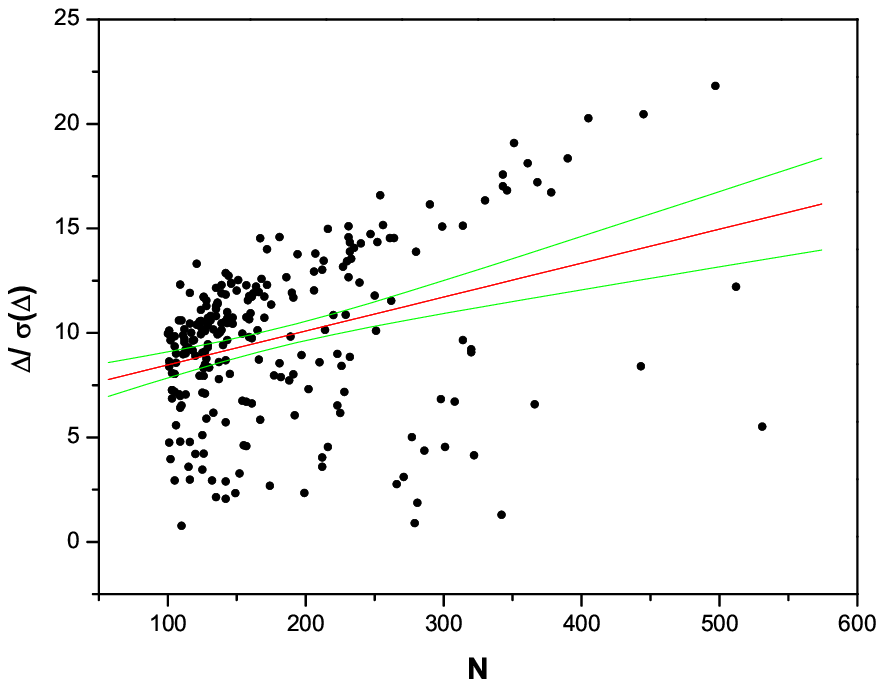}
\caption{The same as in fig.2 but for $\eta$ angles.
Upper panel - equatorial coordinates, bottom panel - supergalactic coordinates.
\label{fig4}}
\end{figure}

\clearpage

\begin{figure}
\epsscale{.30}
\plotone{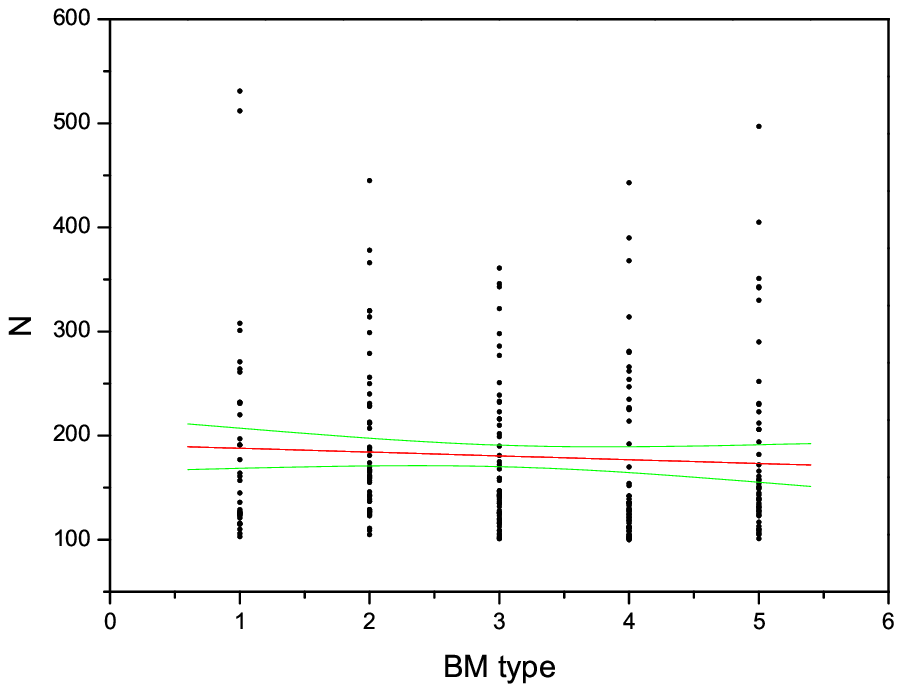}
\plotone{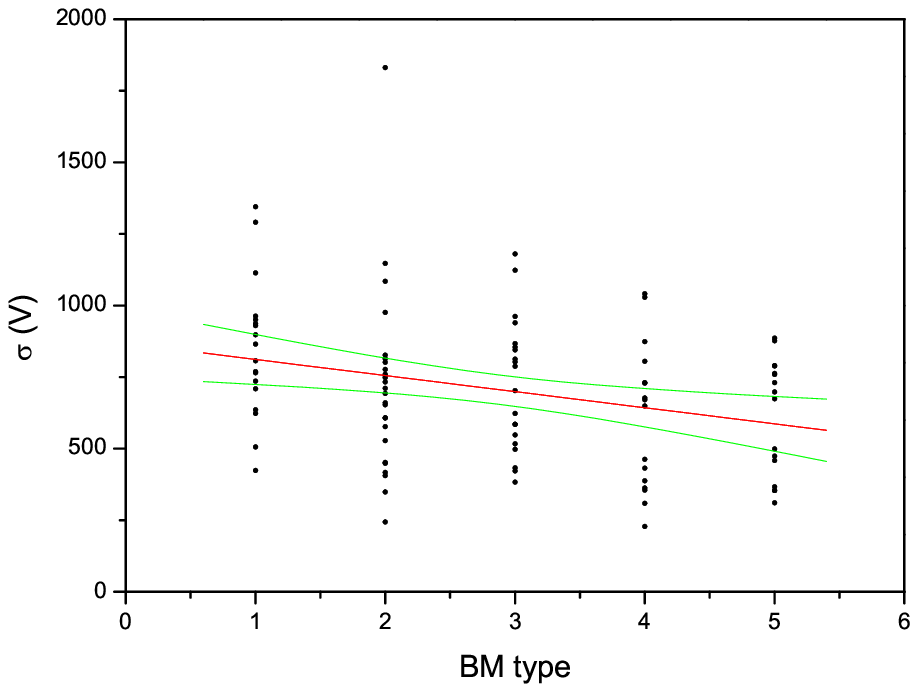}
\plotone{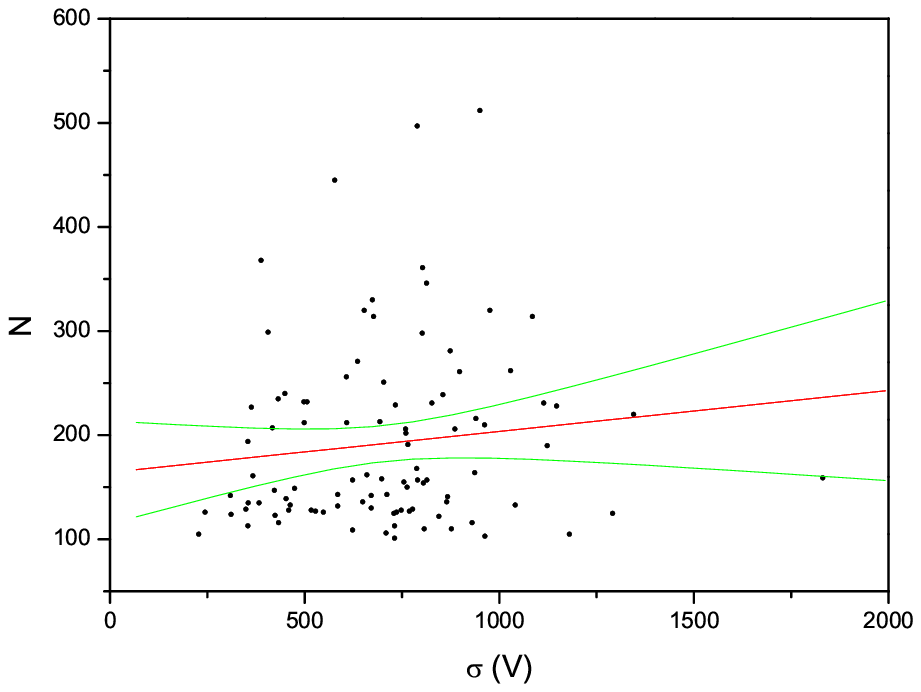}
\caption{The dependence between the
number of galaxies in each cluster $N$ and $BM$ type - left panel,
the velocity dispersion $\sigma(V)$ and $BM$ type - middle panel, the number
of galaxies in each  cluster $N$ and velocity dispersion $\sigma(V)$- right
panel. The bounds error, at confidence level $95\%$, were shown. $BM$ type is
coded as follows: $BMI$ as $1$,  $BMI-II$ as $2$, $BMII$ as $3$,
$BMII-III$ as $4$ and $BMIII$ as $5$.
\label{fig5}}
\end{figure}
 
\clearpage
 
 \begin{table}
\begin{center}
\caption{The results of the linear regression analysis - equatorial coordinates.
\label{Tab.1}}
\begin{tabular}{c|cccc|cccc|cccc}
\tableline\tableline
\multicolumn{1}{c}{}&
\multicolumn{4}{c}{$\chi^2$}&
\multicolumn{4}{c}{$\Delta_1/\sigma(\Delta_1)$}&
\multicolumn{4}{c}{$\Delta/\sigma(\Delta)$}\\
\tableline \tableline
sample&$a$&$\sigma(a)$&$b$&$\sigma(b)$&$a$&$\sigma(a)$&$b$&$\sigma(b)$&$a$&$\sigma(a)$&$b$&$\sigma(b)$\\
\tableline
$N$     &       &        &      &     &        &        &      &      &        &        &      &      \\
$A$     &       &        &      &     &        &        &      &      &        &        &      &      \\
$P$     &$0.025$&$ 0.015$&$34.7$&$1.4$&$0.0018$&$0.0015$&$1.55$&$0.15$&$0.0022$&$0.0015$&$2.08$&$0.14$\\
$\delta$&$0.115$&$ 0.037$&$38.4$&$7.4$&$0.0054$&$0.0021$&$2.60$&$0.41$&$0.0075$&$0.0021$&$3.65$&$0.41$\\
$\eta$  &$0.228$&$ 0.038$&$42.7$&$7.4$&$0.0092$&$0.0024$&$3.46$&$0.49$&$0.0132$&$0.0024$&$4.03$&$0.49$\\
$B$     &       &        &      &     &        &        &      &      &        &        &      &      \\
$P$     &$0.036$&$ 0.021$&$33.8$&$1.6$&$0.0033$&$0.0020$&$1.39$&$0.15$&$0.0032$&$0.0019$&$1.98$&$0.15$\\
$\delta$&$0.178$&$ 0.047$&$25.5$&$7.4$&$0.0096$&$0.0026$&$1.78$&$0.42$&$0.0128$&$0.0028$&$2.58$&$0.43$\\
$\eta$  &$0.324$&$ 0.042$&$26.5$&$6.7$&$0.0174$&$0.0032$&$2.06$&$0.47$&$0.0210$&$0.0030$&$2.69$&$0.47$\\
\tableline
$BM$   &       &        &      &     &        &        &      &      &        &        &      &      \\
$A$     &       &        &      &     &        &        &      &      &        &        &      &      \\
$P$     &$0.085$&$ 0.044$&$36.6$&$1.5$&$0.0137$&$0.0450$&$2.22$&$0.15$&$0.0129$&$0.0409$&$1.42$&$0.13$\\
$\delta$&$-3.98$&$ 4.20 $&$57.0$&$7.4$&$0.0518$&$0.2518$&$3.10$&$0.42$&$-.1572$&$0.2604$&$4.66$&$0.44$\\
$\eta$  &$1.958$&$ 4.979$&$80.6$&$8.4$&$0.1110$&$0.3150$&$4.93$&$0.53$&$0.1122$&$0.3240$&$6.23$&$0.55$\\
$B$     &       &        &      &     &        &        &      &      &        &        &      &      \\
$P$     &$0.272$&$ 0.459$&$35.5$&$1.6$&$0.0868$&$0.0434$&$1.36$&$1.14$&$0.0339$&$0.0427$&$2.10$&$0.14$\\
$\delta$&$-5.26$&$ 4.72 $&$67.3$&$8.0$&$0.0070$&$0.2642$&$3.57$&$0.44$&$-.2328$&$0.2674$&$5.35$&$0.45$\\
$\eta$  &$3.074$&$ 4.320$&$68.6$&$7.2$&$0.2572$&$0.2940$&$4.16$&$0.49$&$0.2294$&$0.3012$&$5.38$&$0.51$\\
\tableline
$\sigma(V)$&    &        &      &     &        &        &      &      &        &        &      &      \\
$A$     &       &        &      &     &        &        &      &      &        &        &      &      \\
$P$     &$0.001$&$ 0.003$&$35.4$&$2.5$&$0.0005$&$0.0003$&$1.13$&$0.22$&$0.0005$&$0.0003$&$1.72$&$0.23$\\
$\delta$&$0.033$&$ 0.024$&$33.4$&$9.1$&$0.0019$&$0.0018$&$2.00$&$0.68$&$0.0014$&$0.0015$&$3.66$&$0.58$\\
$\eta$  &$-.012$&$ 0.041$&$100.$&$16.$&$-.0006$&$0.0026$&$6.19$&$0.98$&$-.0007$&$0.0026$&$7.53$&$0.99$\\
$B$     &       &        &      &     &        &        &      &      &        &        &      &      \\
$P$     &$0.001$&$ 0.003$&$34.9$&$2.4$&$0.0003$&$0.0003$&$1.18$&$0.23$&$0.0002$&$0.0003$&$1.87$&$0.23$\\
$\delta$&$0.013$&$ 0.018$&$31.8$&$6.9$&$0.0007$&$0.0015$&$1.93$&$0.57$&$0.0006$&$0.0014$&$3.20$&$0.52$\\
$\eta$  &$-.019$&$ 0.034$&$85.8$&$13.$&$-.0015$&$0.0024$&$5.61$&$0.89$&$-.0015$&$0.0024$&$6.78$&$0.89$\\
\tableline
\end{tabular}
\end{center}
\end{table}
 
\clearpage

 \begin{table}
\begin{center}
\caption{The results of the linear regression analysis - supergalactic coordinates.
\label{Tab.2}}
\begin{tabular}{c|cccc|cccc|cccc}
\tableline\tableline
\multicolumn{1}{c}{}&
\multicolumn{4}{c}{$\chi^2$}&
\multicolumn{4}{c}{$\Delta_1/\sigma(\Delta_1)$}&
\multicolumn{4}{c}{$\Delta/\sigma(\Delta)$}\\
\tableline \tableline
sample&$a$&$\sigma(a)$&$b$&$\sigma(b)$&$a$&$\sigma(a)$&$b$&$\sigma(b)$&$a$&$\sigma(a)$&$b$&$\sigma(b)$\\
\tableline
$N$     &       &        &      &     &        &        &      &      &        &        &      &      \\
$A$     &       &        &      &     &        &        &      &      &        &        &      &      \\
$P$     &$0.006$&$ 0.016$&$36.2$&$1.5$&$0.0018$&$0.0015$&$1.55$&$0.15$&$0.0020$&$0.0015$&$2.10$&$0.14$\\
$\delta$&$0.280$&$ 0.023$&$ 7.9$&$4.6$&$0.0121$&$0.0017$&$2.06$&$0.34$&$0.0138$&$0.0014$&$2.72$&$0.26$\\
$\eta$  &$0.498$&$ 0.059$&$56.1$&$12.$&$0.0136$&$0.0028$&$5.87$&$0.56$&$0.0162$&$0.0028$&$6.85$&$0.55$\\
$B$     &       &        &      &     &        &        &      &      &        &        &      &      \\
$P$     &$0.031$&$ 0.021$&$34.0$&$1.6$&$0.0035$&$0.0020$&$1.38$&$0.15$&$0.0033$&$0.0019$&$1.99$&$0.14$\\
$\delta$&$0.308$&$ 0.026$&$ 4.5$&$4.0$&$0.0152$&$0.0020$&$1.48$&$0.32$&$0.0178$&$0.0016$&$2.02$&$0.25$\\
$\eta$  &$0.586$&$ 0.064$&$36.4$&$11.$&$0.0228$&$0.0036$&$4.06$&$0.55$&$0.0250$&$0.0034$&$5.03$&$0.54$\\
\tableline
$BM$   &       &        &      &     &        &        &      &      &        &        &      &      \\
$A$     &       &        &      &     &        &        &      &      &        &        &      &      \\
$P$     &$0.049$&$ 0.048$&$36.6$&$1.6$&$0.0775$&$0.0472$&$1.46$&$0.15$&$0.0163$&$0.0448$&$2.22$&$0.15$\\
$\delta$&$5.137$&$ 3.582$&$50.4$&$6.0$&$0.5380$&$0.2286$&$3.40$&$0.38$&$0.3240$&$0.1924$&$4.70$&$0.32$\\
$\eta$  &$19.66$&$ 8.197$&$114.$&$14.$&$0.9366$&$0.3604$&$6.88$&$0.61$&$0.8536$&$0.3618$&$8.45$&$0.61$\\
$B$     &       &        &      &     &        &        &      &      &        &        &      &      \\
$P$     &$0.259$&$ 0.463$&$35.5$&$1.6$&$0.0885$&$0.0435$&$1.36$&$1.15$&$0.0375$&$0.0420$&$2.10$&$0.14$\\
$\delta$&$3.644$&$ 2.941$&$43.3$&$4.9$&$0.5432$&$0.2066$&$2.86$&$0.35$&$0.3382$&$0.1798$&$4.08$&$0.30$\\
$\eta$  &$-10.0$&$ 6.816$&$138.$&$11.$&$-.4794$&$0.2922$&$8.36$&$0.49$&$-.5000$&$0.3006$&$9.27$&$0.51$\\
\tableline
$\sigma(V)$&    &        &      &     &        &        &      &      &        &        &      &      \\
$A$     &       &        &      &     &        &        &      &      &        &        &      &      \\
$P$     &$0.001$&$ 0.004$&$34.6$&$2.7$&$0.0005$&$0.0003$&$1.09$&$0.23$&$0.0004$&$0.0003$&$1.74$&$0.23$\\
$\delta$&$-.028$&$ 0.027$&$66.6$&$10.$&$-.0031$&$0.0018$&$5.25$&$0.69$&$-.0020$&$0.0014$&$5.81$&$0.54$\\
$\eta$  &$-.012$&$ 0.070$&$158.$&$27.$&$-.0006$&$0.0030$&$8.95$&$1.13$&$0.0004$&$0.0030$&$10.2$&$1.14$\\
$B$     &       &        &      &     &        &        &      &      &        &        &      &      \\
$P$     &$0.004$&$ 0.003$&$31.9$&$2.6$&$0.0003$&$0.0003$&$1.17$&$0.23$&$0.0002$&$0.0003$&$1.92$&$0.23$\\
$\delta$&$-.020$&$ 0.022$&$52.4$&$8.4$&$-.0030$&$0.0016$&$4.54$&$0.63$&$-.0022$&$0.0014$&$5.06$&$0.51$\\
$\eta$  &$-.024$&$ 0.059$&$128.$&$22.$&$-.0020$&$0.0028$&$8.06$&$1.05$&$-.0016$&$0.0028$&$9.21$&$1.08$\\
\tableline
\end{tabular}
\end{center}
\end{table}
 
\clearpage
 
\begin{table}
\begin{center}
\caption{The results of the linear regression analysis. We presents  the
dependence between the  number of galaxies in clusters $N$, Bautz-Morgan type $BM$
and the dispersion of the velocities $\sigma(V)$ (fig.4). $P$ - denotes case
of position angle, while $F-G$ Flin-Godlowski method ($\delta$+$\eta$ angles),
respectively.
\label{Tab.3}}
\begin{tabular}{c|cccc}
\tableline\tableline
\multicolumn{1}{c}{}&
\multicolumn{4}{c}{}\\
sample&$a$&$\sigma(a)$&$b$&$\sigma(b)$\\
\tableline
$\sigma(V)$($BM$)&       &        &      &     \\
        &$56.26$&$19.09$&$867.9$&$60.5$\\
\tableline
$N$($BM$)&       &       &      &     \\
sample $A$       &       &       &      &     \\
$F-G$   &$-3.66$&$ 3.94 $&$191.5$&$13.3$\\
$P$     &$-1.92$&$ 1.95 $&$ 93.3$&$ 6.6$\\
sample $B$     &       &        &      &     \\
$F-G$   &$-0.105$&$2.917$&$144.8$&$ 9.8$\\
$P$     &$-0.100$&$0.707$&$ 71.2$&$ 4.8$\\
\tableline
$N$($\sigma(V)$)&       &        &      &     \\
$ sample A$       &       &       &      &     \\
$F-G$   &$0.039$&$0.033$&$164.2$&$24.9$\\
$P$     &$0.021$&$0.016$&$ 79.9$&$12.1$\\
sample $B$       &       &       &      &     \\
$F-G$   &$0.013$&$0.025$&$134.1$&$19.1$\\
$P$     &$0.007$&$0.011$&$ 66.1$&$ 8.6$\\
\tableline
\end{tabular}
\end{center}
\end{table}
 
\clearpage
 
\begin{deluxetable}{lcccccccc}
\tablewidth{0pt}
\tablecaption{The list of investivated clusters}
\tablehead{
\colhead{PF Number}      & \colhead{$\alpha$ [h]}&
\colhead{$\delta$ [deg]} & \colhead{N$_0$}  &
\colhead{N$_1$}          & \colhead{Name}  &
\colhead{$P$}            & \colhead{$\delta_D$}  &
\colhead{$\eta$}}
\startdata
0003-4377 & 0.0332940  & -43.764385 & 252 &  135  & s1173&$\parallel$&$\parallel$&$\parallel$\\
0004-3061 & 0.0424033  & -30.608407 & 125 &   69  & s0001&    0      &$\parallel$&$\parallel$\\
0004-2722 & 0.0496337  & -27.211477 & 162 &   80  & A2716&$\searrow$ &$\parallel$&$\parallel$\\
0009-3469 & 0.0918712  & -34.681737 & 361 &  163  & A2721&    0      &$\parallel$&$\parallel$\\
0016-3529 & 0.1643087  & -35.281307 & 235 &  124  & s0012&    0      &$\parallel$&$\parallel$\\
0017-6446 & 0.1783953  & -64.454778 & 101 &   38  & A2732&    0      &$\parallel$&     0     \\
0020-6640 & 0.2000360  & -66.392536 & 112 &   56  & A2737&    0      &     ?     &$\perp$    \\
0020-4224 & 0.2046051  & -42.236308 & 106 &   52  & A2736&    0      &$\parallel$&$\parallel$\\
0021-1727 & 0.2142782  & -17.267194 & 116 &   33  & s0015&$\parallel$&$\parallel$&$\parallel$\\
0022-1954 & 0.2286907  & -19.534202 & 206 &  106  & A0013&    0      &$\parallel$&$\parallel$\\
0024-4251 & 0.2431575  & -42.500865 & 112 &   54  & A2745&    0      &$\parallel$&$\parallel$\\
0024-2387 & 0.2475143  & -23.860490 & 314 &  169  & A0014&$\parallel$&$\parallel$&$\parallel$\\
0025-2612 & 0.2536075  & -26.112302 & 128 &   73  & A0015&$\parallel$&$\parallel$&$\parallel$\\
0027-3136 & 0.2734246  & -31.358321 & 150 &   82  & A2751&    0      &$\parallel$&$\parallel$\\
0028-4990 & 0.2864895  & -49.896367 & 161 &   76  & A2753&$\parallel$&$\parallel$&$\parallel$\\
0029-3513 & 0.2910976  & -35.125399 & 497 &  230  & A2755&    0      &$\parallel$&$\parallel$\\
0031-2046 & 0.3153715  & -20.456570 & 102 &   53  & s0026&    0      &$\parallel$&$\parallel$\\
0031-4187 & 0.3180634  & -41.863651 & 231 &  101  & A2758&$\searrow$ &$\parallel$&$\parallel$\\
0034-2570 & 0.3429825  & -25.697728 & 213 &  118  & A0022&    0      &$\parallel$&$\parallel$\\
0035-4930 & 0.3514239  & -49.297151 & 168 &   73  & A2764&$\parallel$&$\parallel$&$\parallel$\\
0040-4010 & 0.4030093  & -40.091711 & 129 &   57  & A2771&    0      &$\parallel$&$\parallel$\\
0041-6581 & 0.4136137  & -65.805361 & 139 &   61  & A2770&$\perp$    &$\parallel$&$\perp$    \\
0042-3308 & 0.4259923  & -33.076987 & 172 &   76  & s0041&    0      &$\parallel$&$\parallel$\\
0042-3799 & 0.4273978  & -37.982872 & 160 &   77  & A2772&    0      &$\parallel$&$\parallel$\\
0043-6910 & 0.4301972  & -69.092017 & 172 &   85  & A2775&$\perp$    &     ?     &     ?     \\
0045-5021 & 0.4503612  & -50.207339 & 144 &   58  & A2777&    0      &$\parallel$&$\parallel$\\
0047-2362 & 0.4761155  & -23.614121 & 141 &   76  & A0042&$\parallel$&$\parallel$&$\parallel$\\
0058-3907 & 0.5820089  & -39.062564 & 118 &   67  & s0061&$\parallel$&$\parallel$&     ?     \\
0061-6439 & 0.6129856  & -64.384850 & 101 &   56  & A2796&    0      &     ?     &     ?     \\
0062-2845 & 0.6221154  & -28.448334 & 143 &   66  & A2798&    0      &$\parallel$&$\parallel$\\
0066-2896 & 0.6620048  & -28.953170 & 256 &  108  & A2798&    ?      &$\parallel$&$\parallel$\\
0069-2869 & 0.6950245  & -28.682962 & 445 &  186  & A2804&    0      &$\parallel$&$\parallel$\\
0072-2610 & 0.7234165  & -26.093608 & 194 &   89  & A0088&$\perp$    &$\parallel$&$\parallel$\\
0076-1816 & 0.7601967  & -18.151027 & 135 &   55  & A2816&    0      &$\parallel$&$\parallel$\\
0076-6353 & 0.7650574  & -63.526130 & 299 &  155  & A2819&    0      &$\parallel$&$\perp$    \\
0078-5470 & 0.7843686  & -54.699737 & 145 &   77  & s0077&    ?      &     ?     &     ?     \\
0082-2951 & 0.8211987  & -29.507452 & 110 &   56  & s0084&    0      &$\parallel$&$\parallel$\\
0084-5025 & 0.8476005  & -50.246897 & 118 &   56  & A2827&$\parallel$&$\parallel$&$\parallel$\\
0088-4758 & 0.8873193  & -47.576920 & 113 &   68  & A2836&    0      &     ?     &$\parallel$\\
0089-2178 & 0.8925053  & -21.776334 & 128 &   61  & A0114&    0      &$\parallel$&$\parallel$\\
0092-2637 & 0.9213817  & -26.368269 & 136 &   62  & A0118&    0      &$\parallel$&$\parallel$\\
0095-3094 & 0.9586092  & -30.931841 & 126 &   44  & s0109&$\perp$    &$\parallel$&$\parallel$\\
0097-6680 & 0.9762100  & -66.790348 & 190 &  106  & s0112&    0      &      ?    &$\perp$    \\
0098-3428 & 0.9839121  & -34.279133 & 207 &  103  & A2847&    ?      &$\parallel$&$\parallel$\\
0104-2195 & 1.0481967  & -21.940614 & 157 &   84  & A0133&    0      &$\parallel$&$\parallel$\\
0105-3995 & 1.0520281  & -39.940742 & 264 &  116  & A2857&$\parallel$&$\parallel$&$\parallel$\\
0109-2457 & 1.0975451  & -24.566583 & 136 &   78  & A0141&$\perp$    &$\parallel$&$\parallel$\\
0113-3674 & 1.1319391  & -36.739609 & 116 &   48  & A2871&    0      &$\parallel$&$\parallel$\\
0114-4031 & 1.1455260  & -40.306519 & 164 &   80  & A2874&    0      &$\parallel$&$\parallel$\\
0115-4600 & 1.1567254  & -45.996573 & 261 &  125  & A2877&    ?      &$\parallel$&$\parallel$\\
0117-2480 & 1.1703771  & -24.798778 & 139 &   76  & A0155&    0      &$\parallel$&$\parallel$\\
0132-6468 & 1.3204981  & -64.674528 & 111 &   49  & A2899&$\parallel$&$\parallel$&      ?    \\
0133-2109 & 1.3305868  & -21.083654 & 131 &   69  & A0177&    0      &$\parallel$&$\parallel$\\
0145-5012 & 1.4548714  & -50.110378 & 247 &  104  & A2912&    0      &$\parallel$&$\parallel$\\
0153-2483 & 1.5302729  & -24.829812 & 186 &   87  & A2921&    0      &$\parallel$&$\parallel$\\
0154-2697 & 1.5492938  & -26.963567 & 135 &   68  & A2924&    0      &$\parallel$&$\parallel$\\
0156-2609 & 1.5607925  & -26.080661 & 240 &  112  & A0214&    ?      &$\parallel$&$\parallel$\\
0157-3273 & 1.5709466  & -32.727195 & 127 &   61  & s0167&    0      &$\parallel$&$\parallel$\\
0157-2751 & 1.5755244  & -27.507008 & 170 &   91  & A2928&$\searrow$ &$\parallel$&$\parallel$\\
0168-5458 & 1.6854490  & -54.572596 & 206 &   95  & A2933&$\perp$    &$\parallel$&$\parallel$\\
0175-5305 & 1.7507039  & -53.046656 & 131 &   53  & A2941&    0      &     ?     &$\parallel$\\
0179-6283 & 1.7986149  & -62.827489 & 111 &   43  & s0194&    0      &$\parallel$&$\searrow$ \\
0184-2698 & 1.8463826  & -26.976824 & 117 &   41  & A2945&    0      &$\parallel$&$\parallel$\\
0202-4477 & 2.0235875  & -44.764599 & 127 &   66  & s0217&    0      &     ?     &$\parallel$\\
0206-4119 & 2.0636706  & -41.180976 & 136 &   63  & A2969&$\perp$    &     ?     &$\parallel$\\
0207-3573 & 2.0796464  & -35.728896 & 105 &   57  & A2970&    0      &     ?     &$\parallel$\\
0209-2715 & 2.0957319  & -27.140202 & 129 &   64  & A2972&$\parallel$&$\parallel$&$\parallel$\\
0211-2615 & 2.1148005  & -26.147306 & 112 &   53  & A2973&    0      &     ?     &$\parallel$\\
0221-2527 & 2.2131866  & -25.266329 & 126 &   60  & A0325&$\parallel$&$\parallel$&$\parallel$\\
0221-4723 & 2.2145944  & -47.226227 & 161 &   79  & A2988&    0      &$\parallel$&$\parallel$\\
0221-3443 & 2.2183616  & -34.422916 & 103 &   50  & s0233&    0      &     0     &$\parallel$\\
0233-1903 & 2.3331613  & -19.023636 & 119 &   61  & A3005&$\parallel$&$\parallel$&$\parallel$\\
0257-5948 & 2.5765992  & -59.475835 & 189 &  100  & s0280&    0      &     ?     &$\searrow$ \\
0257-3359 & 2.5785227  & -33.585227 & 113 &   66  & s0278&$\perp$    &$\perp$    &$\parallel$\\
0258-5925 & 2.5855871  & -59.241983 & 108 &   62  & s0284&    0      &     0     &     0     \\
0260-1938 & 2.6096109  & -19.372846 & 103 &   47  & A0367&$\searrow$ &$\parallel$&$\parallel$\\
0263-5237 & 2.6360181  & -52.368330 & 123 &   67  & A3038&$\perp$    &$\perp$    &     ?     \\
0270-2864 & 2.7005975  & -28.638252 & 166 &   84  & A3041&    0      &$\parallel$&$\parallel$\\
0273-2634 & 2.7378078  & -26.338302 & 251 &  112  & A0380&$\parallel$&$\parallel$&$\parallel$\\
0275-4645 & 2.7581401  & -46.444123 & 197 &  101  & A3034&    0      &     0     &$\parallel$\\
0284-7138 & 2.8401619  & -71.376999 & 143 &   81  & s0303&$\perp$    &     0     &$\perp$    \\
0284-4629 & 2.8422326  & -46.289948 & 137 &   60  & A3059&    ?      &     0     &$\parallel$\\
0285-2494 & 2.8567712  & -24.932547 & 191 &   98  & A0389&    0      &$\parallel$&$\parallel$\\
0286-2549 & 2.8684629  & -25.483217 & 262 &  120  & A3062&    0      &$\parallel$&$\parallel$\\
0293-2261 & 2.9300204  & -22.600655 & 188 &   84  & A3069&    0      &$\searrow$ &$\parallel$\\
0296-5278 & 2.9631675  & -52.774388 & 109 &   59  & A3074&    0      &$\searrow$ &$\parallel$\\
0299-5183 & 2.9982249  & -51.828647 & 232 &  103  & A3078&$\parallel$&     ?     &$\parallel$\\
0304-7930 & 3.0468766  & -79.293804 & 115 &   59  & s0322&    0      &$\parallel$&$\perp$    \\
0313-2364 & 3.1387755  & -23.639519 & 228 &  118  & A0419&$\parallel$&$\searrow$ &$\parallel$\\
0322-3835 & 3.2280135  & -38.341075 & 106 &   55  & A3098&    0      &     0     &     ?     \\
0325-4274 & 3.2580363  & -42.733818 & 109 &   38  & A3107&    0      &$\searrow$ &     ?     \\
0327-5091 & 3.2741445  & -50.901538 & 128 &   66  & A3110&    0      &     0     &$\parallel$\\
0328-4650 & 3.2881271  & -46.494930 & 181 &   75  & s0335&$\searrow$ &$\searrow$ &$\parallel$\\
0329-4427 & 3.2953835  & -44.261528 & 512 &  236  & A3112&$\perp$    &$\searrow$ &$\parallel$\\
0331-5398 & 3.3183956  & -53.979633 & 154 &   85  & s0339&$\perp$    &$\searrow$ &$\parallel$\\
0332-7192 & 3.3204402  & -71.918233 & 120 &   71  & A3117&    0      &$\parallel$&$\perp$    \\
0334-4314 & 3.3469023  & -43.136727 & 137 &   59  & s0343&    0      &$\searrow$ &$\parallel$\\
0336-4135 & 3.3666497  & -41.341134 & 155 &   82  & A3122&    0      &     0     &     ?     \\
0336-4554 & 3.3681819  & -45.536097 & 106 &   37  & s0345&    0      &$\parallel$&$\parallel$\\
0347-5571 & 3.4783319  & -55.709845 & 133 &   63  & A3126&    0      &$\searrow$ &$\searrow$ \\
0348-4604 & 3.4821406  & -46.031863 & 127 &   49  & s0356&    0      &$\perp$    &$\parallel$\\
0350-5258 & 3.5096315  & -52.578089 & 298 &  165  & A3128&    0      &$\searrow$ &     ?     \\
0353-7192 & 3.5393146  & -71.910959 & 250 &  126  & A3136&    ?      &$\parallel$&$\perp$    \\
0360-4072 & 3.6032229  & -40.717086 & 109 &   50  & A3140&    0      &$\perp$    &$\parallel$\\
0361-3977 & 3.6157370  & -39.769377 & 157 &   65  & A3142&    0      &$\perp$    &$\parallel$\\
0364-3272 & 3.6495967  & -32.718875 & 125 &   53  & A3148&    0      &$\perp$    &     ?     \\
0370-5364 & 3.7097584  & -53.639887 & 320 &  146  & A3158&    0      &     0     &$\parallel$\\
0376-2427 & 3.7641976  & -24.265911 & 126 &   56  & A0458&    0      &$\perp$    &$\parallel$\\
0380-4555 & 3.8023921  & -45.541180 & 115 &   56  & s0393&    0      &     0     &     0     \\
0380-3349 & 3.8076228  & -33.483048 & 167 &   70  & A3169&$\perp$    &$\perp$    &$\parallel$\\
0381-1789 & 3.8171204  & -17.885726 & 308 &  136  & A0464&$\perp$    &$\perp$    &$\parallel$\\
0383-7395 & 3.8395870  & -73.944664 & 165 &   88  & A3186&$\searrow$ &$\parallel$&$\perp$    \\
0395-6285 & 3.9539293  & -62.849691 & 103 &   48  & A3191&    0      &     0     &     0     \\
0396-2698 & 3.9698326  & -26.970222 & 149 &   79  & A3188&    0      &$\perp$    &     0     \\
0398-3012 & 3.9822735  & -30.113240 & 157 &   81  & A3194&$\searrow$ &$\perp$    &$\parallel$\\
0400-5366 & 4.0014897  & -53.658246 & 116 &   56  & A3202&$\searrow$ &$\searrow$ &     0     \\
0406-1733 & 4.0672780  & -17.326625 & 125 &   58  & A0473&    ?      &$\perp$    &$\parallel$\\
0407-6535 & 4.0720651  & -65.344262 & 214 &  100  & A3216&$\perp$    &$\searrow$ &$\perp$    \\
0407-7500 & 4.0758095  & -74.994037 & 103 &   50  & s0428&    0      &     ?     &     ?     \\
0407-4397 & 4.0798443  & -43.966323 & 105 &   62  & s0416&    0      &$\perp$    &$\parallel$\\
0413-3091 & 4.1336041  & -30.907580 & 271 &  131  & A3223&    0      &$\perp$    &     0     \\
0418-6383 & 4.1887482  & -63.829294 & 126 &   54  & A3230&    0      &$\searrow$ &     ?     \\
0423-4564 & 4.2388139  & -45.635285 & 223 &  106  & A3235&$\perp$    &$\perp$    &     ?     \\
0430-4434 & 4.3097327  & -44.333642 & 102 &   49  & s0437&$\perp$    &$\perp$    &     0     \\
0431-4515 & 4.3154473  & -45.149298 & 225 &   90  & A3245&$\perp$    &$\perp$    &     ?     \\
0436-4653 & 4.3676988  & -46.520496 & 443 &  200  & A3247&$\perp$    &$\perp$    &     0     \\
0437-2443 & 4.3765635  & -24.420645 & 142 &   71  & A0487&    0      &$\perp$    &     0     \\
0442-3619 & 4.4243921  & -36.181623 & 116 &   54  & A3253&    0      &$\perp$    &     0     \\
0448-6701 & 4.4847770  & -67.008555 & 125 &   70  & s0467&    0      &     0     &     0     \\
0449-5374 & 4.4945515  & -53.737334 & 212 &  122  & s0463&$\perp$    &$\perp$    &     0     \\
0451-4923 & 4.5144323  & -49.229827 & 120 &   52  & A3264&$\parallel$&     ?     &     0     \\
0451-6138 & 4.5173059  & -61.371014 & 314 &  144  & A3266&    0      &$\searrow$ &$\perp$    \\
0451-4625 & 4.5187527  & -46.243585 & 286 &  151  & s0468&    0      &$\perp$    &     0     \\
0453-4613 & 4.5377246  & -46.125642 & 322 &  163  & s0468&    0      &$\perp$    &     0     \\
0454-3268 & 4.5478622  & -32.679840 & 199 &   85  & A3269&    0      &$\perp$    &     0     \\
0458-3576 & 4.5827851  & -35.756298 & 266 &  129  & A3277&    0      &$\perp$    &     0     \\
0462-2040 & 4.6221526  & -20.393051 & 174 &   74  & A0499&    0      &$\perp$    &     0     \\
0465-2208 & 4.6516903  & -22.078465 & 152 &   61  & A0500&$\perp$    &$\perp$    &     0     \\
0471-3288 & 4.7141684  & -32.873363 & 142 &   73  & s0491&    0      &$\perp$    &     0     \\
0471-4503 & 4.7151981  & -45.026389 & 531 &  253  & A3284&    0      &$\perp$    &$\searrow$ \\
0476-2550 & 4.7691067  & -25.496414 & 342 &  167  & A0511&    0      &$\perp$    &     0     \\
0480-2050 & 4.8004748  & -20.495970 & 281 &  153  & A0514&    0      &$\perp$    &     0     \\
0497-3012 & 4.9703154  & -30.119282 & 110 &   44  & A3297&    0      &$\perp$    &     0     \\
0500-3868 & 5.0092522  & -38.670735 & 301 &  148  & A3301&$\perp$    &$\perp$    &     0     \\
0506-5695 & 5.0694330  & -56.942871 & 192 &   80  & A3312&    0      &     ?     &$\parallel$\\
0521-4180 & 5.2189798  & -41.790715 & 212 &  105  & s0515&$\searrow$ &$\perp$    &$\searrow$ \\
0524-4910 & 5.2496234  & -49.090224 & 216 &  103  & A3330&$\perp$    &$\perp$    &     0     \\
0542-3150 & 5.4274828  & -31.497747 & 132 &   66  & A3341&$\perp$    &$\perp$    &     ?     \\
0542-3070 & 5.4275138  & -30.696179 & 279 &  155  & A3342&    0      &$\perp$    &     0     \\
0557-2851 & 5.5729561  & -28.507059 & 135 &   70  & A3354&    0      &$\perp$    &     0     \\
2036-5287 & 20.3650881 & -52.867343 & 128 &   60  & A3675&$\parallel$&$\parallel$&$\perp$    \\
2053-6310 & 20.5383289 & -63.096072 & 230 &  139  & A3687&$\parallel$&$\parallel$&$\perp$    \\
2070-3523 & 20.7036454 & -35.229947 & 110 &   53  & A3705&    0      &$\searrow$ &     0     \\
2077-3251 & 20.7745307 & -32.504550 & 161 &   81  & A3712&    0      &     0     &$\searrow$ \\
2086-5271 & 20.8646808 & -52.702585 & 229 &  148  & A3716&    0      &$\parallel$&$\perp$    \\
2092-5487 & 20.9281238 & -54.863643 & 126 &   63  & A3718&$\perp$    &$\parallel$&$\perp$    \\
2098-3648 & 20.9878184 & -36.470935 & 105 &   39  & A3727&    0      &     ?     &$\searrow$ \\
2102-2808 & 21.0272287 & -28.071446 & 202 &  111  & A3733&$\parallel$&     0     &$\parallel$\\
2109-3878 & 21.0992361 & -38.778351 & 191 &   92  & A3740&$\parallel$&$\parallel$&$\searrow$ \\
2112-3960 & 21.1207422 & -39.594784 & 177 &  109  & s0922&    0      &$\parallel$&$\searrow$ \\
2130-4535 & 21.3091848 & -45.347944 & 170 &   83  & A3757&    0      &     0     &$\perp$    \\
2143-3473 & 21.4362913 & -34.726841 & 130 &   57  & A3764&    0      &     ?     &     0     \\
2143-3477 & 21.4376594 & -34.762871 & 142 &   62  & A3764&    0      &     ?     &$\searrow$ \\
2145-4267 & 21.4512883 & -42.662478 & 105 &   48  & A3767&    0      &     ?     &     0     \\
2149-5088 & 21.4923816 & -50.872873 & 143 &   75  & A3771&    0      &$\parallel$&$\perp$    \\
2153-4327 & 21.5352712 & -43.268813 & 159 &   72  & A3775&    0      &$\parallel$&$\searrow$ \\
2154-2275 & 21.5464792 & -22.745785 & 226 &  119  & A3778&$\parallel$&     ?     &     ?     \\
2157-5359 & 21.5776938 & -53.581931 & 175 &  109  & A3758&    0      &$\parallel$&$\perp$    \\
2158-6206 & 21.5821533 & -62.055750 & 233 &  121  & A3782&    0      &$\parallel$&$\perp$    \\
2160-2326 & 21.6092861 & -23.258903 & 145 &   72  & A2357&$\parallel$&     0     &     ?     \\
2165-5159 & 21.6564705 & -51.588395 & 378 &  208  & A3976&$\parallel$&$\parallel$&$\perp$    \\
2172-3894 & 21.7226966 & -38.933406 & 137 &   66  & s0964&    ?      &$\parallel$&$\searrow$ \\
2172-1868 & 21.7243179 & -18.677874 & 101 &   62  & A2365&    0      &$\searrow$ &     ?     \\
2175-2414 & 21.7530741 & -24.132785 & 182 &   97  & A2371&$\parallel$&     ?     &     ?     \\
2175-2617 & 21.7574597 & -26.166827 & 223 &  115  & A3805&$\parallel$&$\parallel$&$\parallel$\\
2176-5163 & 21.7635527 & -51.622880 & 123 &   53  & s0968&$\parallel$&$\parallel$&$\searrow$ \\
2177-4386 & 21.7722038 & -43.856255 & 212 &  118  & A3809&    0      &$\parallel$&$\searrow$ \\
2177-5727 & 21.7788819 & -57.267994 & 346 &  187  & A3806&    0      &$\parallel$&$\perp$    \\
2179-4598 & 21.7948239 & -45.973743 & 280 &  141  & s0974&$\parallel$&$\parallel$&$\searrow$ \\
2181-3068 & 21.8158351 & -30.675021 & 210 &  116  & A3814&    0      &$\parallel$&$\parallel$\\
2184-5538 & 21.8460586 & -55.374763 & 139 &   75  & A3816&    0      &$\parallel$&$\perp$    \\
2187-1958 & 21.8725458 & -19.573461 & 142 &   79  & A2384&    0      &     0     &     ?     \\
2190-5774 & 21.9037001 & -57.738107 & 698 &  385  & A3822&    0      &$\parallel$&$\perp$    \\
2197-6040 & 21.9788027 & -60.394643 & 158 &   82  & A3825&    0      &$\parallel$&$\perp$    \\
2201-6666 & 22.0138069 & -66.654358 & 135 &   82  & s0984&    0      &     ?     &$\perp$    \\
2202-6002 & 22.0276285 & -60.018229 & 231 &  105  & A3827&$\parallel$&$\parallel$&$\perp$    \\
2217-5177 & 22.1707167 & -51.766387 & 129 &   71  & A3836&    0      &$\parallel$&$\searrow$ \\
2222-3471 & 22.2295665 & -34.706016 & 101 &   44  & A3844&    0      &$\parallel$&     ?     \\
2223-3675 & 22.2390700 & -36.741350 & 147 &   66  & s1005&    0      &$\parallel$&$\parallel$\\
2225-5153 & 22.2589927 & -51.522674 & 100 &   52  & A3849&$\searrow$ &$\parallel$&$\searrow$ \\
2229-3570 & 22.2937967 & -35.691106 & 105 &   53  & A3854&    ?      &     ?     &$\parallel$\\
2230-3890 & 22.3097047 & -38.892979 & 125 &   54  & A3856&$\parallel$&$\parallel$&$\parallel$\\
2233-4598 & 22.3312483 & -45.978565 & 142 &   55  & A3862&$\perp$    &$\parallel$&     0     \\
2234-5249 & 22.3418251 & -52.489333 & 216 &   73  & A3864&    0      &$\parallel$&     0     \\
2241-6428 & 22.4175145 & -64.274067 & 123 &   70  & s1022&    0      &$\parallel$&$\perp$    \\
2243-4774 & 22.4361028 & -47.732841 & 121 &   61  & A3876&    0      &$\parallel$&$\searrow$ \\
2243-5723 & 22.4374814 & -57.221621 & 124 &   65  & A3875&$\perp$    &$\parallel$&     ?     \\
2244-8018 & 22.4465719 & -80.176807 & 159 &  102  & s1014&    0      &$\parallel$&$\perp$    \\
2244-5592 & 22.4487734 & -55.918123 & 351 &  152  & s1023&$\perp$    &$\parallel$&$\perp$    \\
2245-3049 & 22.4546410 & -30.484338 & 239 &  121  & A3880&    0      &$\parallel$&$\parallel$\\
2249-4810 & 22.4930878 & -48.098028 & 157 &   79  & A3883&    0      &$\parallel$&$\searrow$ \\
2251-5468 & 22.5191493 & -54.678874 & 405 &  203  & A3886&$\perp$    &$\parallel$&$\perp$    \\
2256-3778 & 22.5697146 & -37.772273 & 159 &   89  & A3888&    0      &$\parallel$&$\parallel$\\
2260-2448 & 22.6098915 & -24.470898 & 220 &  115  & s1043&$\parallel$&$\parallel$&$\parallel$\\
2265-3654 & 22.6511151 & -36.538601 & 368 &  168  & A3895&    0      &$\parallel$&$\parallel$\\
2277-5266 & 22.7728262 & -52.659033 & 254 &  109  & A3911&$\searrow$ &$\parallel$&$\searrow$ \\
2279-7177 & 22.7951191 & -71.761058 & 142 &   70  & A3916&$\parallel$&$\parallel$&$\perp$    \\
2282-6440 & 22.8261247 & -64.392424 & 143 &   72  & A3921&    0      &$\parallel$&$\perp$    \\
2284-4641 & 22.8467747 & -46.406595 & 390 &  188  & s1066&    0      &$\parallel$&$\parallel$\\
2286-3346 & 22.8653802 & -33.457812 & 108 &   41  & A3926&    0      &$\parallel$&$\parallel$\\
2290-5820 & 22.9037119 & -58.196725 & 124 &   69  & A3939&    0      &$\parallel$&$\searrow$ \\
2292-5564 & 22.9209509 & -55.634206 & 146 &   69  & A3938&    0      &$\parallel$&     0     \\
2297-3077 & 22.9794554 & -30.769579 & 133 &   68  & s1075&    0      &$\parallel$&$\parallel$\\
2299-5613 & 22.9932807 & -56.122212 & 128 &   55  & A3950&    ?      &$\parallel$&$\searrow$ \\
2310-4524 & 23.1013885 & -45.234819 & 126 &   54  & A3970&$\parallel$&$\parallel$&$\parallel$\\
2314-1991 & 23.1444582 & -19.901091 & 154 &   79  & A2538&    ?      &$\parallel$&$\parallel$\\
2317-2446 & 23.1707661 & -24.452256 & 127 &   65  & A2542&    ?      &$\parallel$&$\parallel$\\
2317-2276 & 23.1767610 & -22.754556 & 166 &   65  & A2546&    0      &$\parallel$&$\parallel$\\
2318-2054 & 23.1897451 & -20.532686 & 140 &   66  & A2548&    0      &$\parallel$&$\parallel$\\
2319-2888 & 23.1935895 & -28.876656 & 227 &  119  & A3978&    0      &$\parallel$&$\parallel$\\
2320-2162 & 23.2017030 & -21.613963 & 231 &  116  & A2554&    0      &$\parallel$&$\parallel$\\
2320-7329 & 23.2070355 & -73.283348 & 121 &   39  & s1097&$\perp$    &$\parallel$&$\perp$    \\
2323-4265 & 23.2334892 & -42.648829 & 135 &   63  & s1101&    0      &$\parallel$&$\parallel$\\
2325-3781 & 23.2561310 & -37.809397 & 129 &   58  & A3984&$\parallel$&$\parallel$&$\parallel$\\
2328-7492 & 23.2873598 & -74.915989 & 181 &   60  & s1105&    0      &$\parallel$&$\perp$    \\
2331-4225 & 23.3110602 & -42.249813 & 343 &  163  & s1111&    0      &$\parallel$&$\parallel$\\
2334-7330 & 23.3424547 & -73.295282 & 146 &   57  & A3992&$\parallel$&$\parallel$&$\perp$    \\
2335-2314 & 23.3528905 & -23.131756 & 232 &  117  & A2580&    0      &$\parallel$&$\parallel$\\
2335-4186 & 23.3583306 & -41.854040 & 124 &   62  & A3998&    0      &$\parallel$&$\parallel$\\
2337-2046 & 23.3767020 & -20.457046 & 138 &   73  & A2583&    0      &$\parallel$&$\parallel$\\
2338-5170 & 23.3838174 & -51.697333 & 139 &   61  & A3999&    0      &$\parallel$&$\parallel$\\
2346-6242 & 23.4609620 & -62.415941 & 109 &   37  & A4006&$\perp$    &$\parallel$&      ?    \\
2350-3920 & 23.5033142 & -39.196389 & 123 &   67  & A4008&    ?      &$\parallel$&$\parallel$\\
2359-5228 & 23.5921173 & -52.279494 & 124 &   56  & A4018&    0      &$\parallel$&$\parallel$\\
2375-2604 & 23.7545938 & -26.033124 & 122 &   68  & A2660&$\searrow$ &$\parallel$&$\parallel$\\
2387-3433 & 23.8753332 & -34.322018 & 232 &  134  & s1157&    0      &$\parallel$&$\parallel$\\
2391-2767 & 23.9149501 & -27.661862 & 113 &   58  & A4053&    0      &$\parallel$&$\parallel$\\
2391-2044 & 23.9196846 & -20.430089 & 151 &   69  & A2679&    0      &$\parallel$&$\parallel$\\
2393-2108 & 23.9385755 & -21.070632 & 101 &   49  & A2680&    ?      &$\parallel$&$\parallel$\\
0320-2704 &  3.2014478 & -27.032059 & 320 &  150  & A3094&    0      &     ?     &$\parallel$\\
0383-1801 &  3.8305438 & -18.009478 & 366 &  172  & A3175&$\perp$    &$\perp$    &$\parallel$\\
0468-4526 &  4.6835843 & -45.252794 & 277 &  126  & s0486&    0      &$\perp$    &$\searrow$ \\
2020-5671 & 20.2003641 & -56.706966 & 147 &   73  & s0854&    0      &$\parallel$&$\perp$    \\
2095-4683 & 20.9545182 & -46.826834 & 343 &  132  & A3720&    0      &$\parallel$&$\perp$    \\
2265-1738 & 22.6542356 & -17.370522 & 126 &   54  & A3897&    0      &$\parallel$&$\parallel$\\
2308-4429 & 23.0869996 & -44.289679 & 290 &  133  & A3969&$\parallel$&$\parallel$&$\parallel$\\
2332-7391 & 23.3291361 & -73.904396 & 167 &   52  & s1104&    0      &$\parallel$&$\perp$    \\
2378-2816 & 23.7840810 & -28.151115 & 330 &  169  & A4037&    0      &$\parallel$&$\perp$    \\
\enddata
\end{deluxetable}
 
\end{document}